\documentclass[pra,twocolumn,superscriptaddress]{revtex4-1}
\usepackage[english]{babel}
\usepackage{graphicx}
\usepackage{amsmath}
\usepackage{amssymb}
\usepackage{amsfonts}
\usepackage{color}
\usepackage{graphicx}
\usepackage{amsfonts}
\usepackage{textcomp}
\usepackage[normalem]{ulem}
\newcommand{\beq}{\begin{equation}}
\newcommand{\eeq}{\end{equation}}
\newcommand{\bea}{\begin{eqnarray}}
\newcommand{\eea}{\end{eqnarray}}

\begin{document}

\title{Measurements of the magnetic properties of conduction electrons}

\author{V.~M.~Pudalov$^{1,2}$\\
$^1$P.~N.~Lebedev Physical Institute,  Moscow 119991, Russia.\\
$^2$National Research University Higher School of Economics, \\Moscow 101000, Russia
}

\begin{abstract}
We consider various methods and techniques for measuring electron magnetization and susceptibility, which are used in experimental condensed matter physics. The list of considered methods for macroscopic measurements includes magnetomechanic, electromagnetic, modulation-type, and also thermodynamic methods based on the chemical potential variation measurements. We also consider local methods of magnetic measurements based on the spin Hall effects, NV-centers. Several scanning probe magnetometers-microscopes are considered, such as magnetic resonance force microscope, SQUID-microscope, and Hall microscope. The review focuses on the spin magnetization measurements of electrons in non-magnetic materials and artificial systems, particularly, in low-dimensional electron systems in semiconductors and in nanosystems, which came to the forefront in recent years.
\end{abstract}
\maketitle

\tableofcontents

\section{Introduction}
Due to the additivity of thermodynamic
quantities, measurements of any of them become challenging
 with reducing the size of the studied
sample. Experimentalists  face with this
problem even when dealing with small three-dimensional
objects such as, for example,
``whiskers'' of Zn, Bi, Sn, whose typical sizes
 are $1\times 10\times (100 - 1000)\mu$m$^3$.

Indeed, the magnetic susceptibility of nonmagnetic metals is typically
 $\sim 10^{-6}$, therefore, the magnetization
of such a sample in a field of $10^3$Oe is $\sim 10^{-12}$\,CGS,
that is several orders of magnitude less than the sensitivity threshold
of traditional laboratory magnetometers - torsion-  \cite{griessen_1973, vandrkooy_1969}, Faraday- (or ``magnetic balance''), vibration-type (so-called, Foner-magnetometer) \cite{foner_1959}, and others.

For such small samples, the change in magnetic
susceptibility by 1\% in a field of $B = 10^3$Oe will cause
a change in the magnetic flux through the
sample crossection $1\times 10\mu$m$^2$ of about  $10^{-12}$\,CGS, which
is approximately $10^{-4}$ flux quantum and also lies beyond
the sensitivity threshold of SQUID magnetometers.

This seemingly purely technical problem for a long time
remained an obstacle to studying
magnetic properties of two-dimensional (2D) electron systems
 in which the effective thickness of the electron layer
is of the order of the Fermi wavelength $(10 - 50$)\AA
and where the typical number of electrons in the sample
is total $10^8 - 10^9$.

The characteristic energy associated with the
sought for changes in magnetic properties are not so
small, $\mu_B B \sim 0.1$\,meV  per electron. This obviously
means that the difficulty of measuring the magnetic properties of two-dimensional
and ultrathin samples are associated not with a smallness  of
the effects, but with the inadequacy of traditional
methods for measuring the properties of samples of small thicknesses.

Clearly,  to overcome the problem,
different measurement methods are needed  in which the
signal magnitude does not decrease  proportionally  to
the sample volume. This review addresses a number of
such, in fact, classical methods that were successfully
used in practice.

In the field of  condensed matter magnetometry  there is a review
by Usher and Elliot \cite{usher09},  which
considers classical methods for measuring orbital electron magnetization, and their application
for studying the quantum Hall effect and
related phenomena. There are also a number of monographs (eg, \cite{chechernikov_book}), where
techniques are considered for measurements with ferromagnetic materials.
In this review, unlike \cite{usher09, chechernikov_book}, the  focus is
on the methods for measuring {\em spin} rather than orbital  magnetization
of electrons in non-magnetic materials, the former is
usually much less than the orbital one.

For completeness and reader convenience
this review also briefly mentioned not only the spin magnetism,
but also the orbital magnetism of electronic systems,
including those that have already been described in \cite{usher09}.
However, we supplement this description with some
results omitted in \cite{usher09}.

Further, the review describes more modern methods developed in the last 20 years, in connection with
with the task of studying the spin properties of strongly correlated electrons in low-dimensional
systems. Recently,  thermodynamic methods turned out to be  among the most fruitful; they
are based on  measurement of chemical potential derivatives for the two-dimensional systems.
The consideration is accompanied by a description of several key
physical results obtained by these methods.

Finally, the review considers local methods of measurements, including various types of scanning magnetic microscopes,  booming recently in connection with the numerous tasks of spintronics, manipulating
with single spins, biophysics, and virology.

\section{Traditional methods of electron magnetization measurements}

\subsection{Electromechanic methods
}

These methods can be divided into two classes:\\
(a) based on the measurements of the force acting on a
sample in an inhomogeneous magnetic field $\textbf{F}=(\textbf{M}
  \nabla)   \textbf{B} $
(Faraday magnetometer), or torque $\textbf{L}= [\textbf{M}\times \textbf{B}]$ - in case of an anisotropic sample in a uniform  magnetic field (torsion magnetometer) and \\
(b)  based on  electromagnetic
induction measurements (Foner magnetometer).

\subsubsection{Torsion magnetometer}

This type of magnetometer is based on ``torsion  balance'' introduced in
the everyday use of experimental physics
 at the end of the 18th century by C.-A. de Coulomb to measure electrical
forces, and by  H. Cavendish -
for measuring gravitational forces. In the contemporary experimental
 physics, laboratory magnetometers
are ubiquitous for measuring
in a uniform field $\textbf{B}$ the torque acting on an anisotropic sample.
In the torsion magnetometers, this torque is compensated by forces from
an elastic element deformation.

\subsubsection{Torsion magnetometers with electric detection
}

\begin{figure}[h]
\begin{center}
\includegraphics[width=200pt]{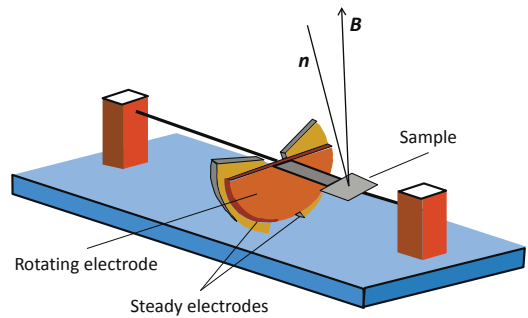}
\caption{Schematic design of the torsion magnetometer from Ref.~\cite{eisenstein_APL_1985}.
$\textbf{n}$ -- normal to the sample plane, $\textbf{B}$-- magnetic field vector.}
\label{fig:eisenstein-torsion}
\end{center}
\end{figure}

To measure the deformation of an elastic element,  capacitive, inductive, or optical
sensors are used. Capacitive deformation sensors \cite{griessen_1973}
starting from the 1960s to the present were successfully used
 for measurements of the oscillatory magnetization (de Haas-van Alphen effect, dHvA).
In regard to the problem of measuring  magnetic properties of low-dimension systems,
the  torsion balance scales were adapted by Eisenstein et al. to measure the
dHvA effect for electrons in a 2D system \cite{eisenstein_APL_1985, eisenstein_PRL_1985}; design
of these scales is shown schematically in Fig.~\ref{fig:eisenstein-torsion}. The sample
 - GaAs-AlGaAs heterostructure - with 2D electron
gas is attached to a thin elastic thread (Pt-W,
with a diameter of 37$\mu$m and 2 cm long), stretched
perpendicular to the magnetic field direction. The orbital magnetic moment of electrons in 2D
systems  ${\overline{M}}$  is the  partial derivative of the free energy with
respect to magnetic  field:

$$
\overline{M}=-\left[{\frac{\partial F}{\partial \overline{B}}}\right]_N.
$$

For isotropic samples (ignoring geometric demagnetizing factor), the  field-induced magnetic moment
 $\overline{M}$  is parallel to  $\overline{B}$ and, therefore, the torque does not arise. For a 2D electron system
the  induced orbital moment is always directed normal to the 2D plane, due to
cyclotron motion in the 2D plane. This magnetic moment causes a mechanical
torque acting on the sample
$$
\textbf{L}=\textbf{M}\times \textbf{B} + \overline{d}\times\nabla(\overline{M}\cdot \overline{B}),
$$
where the second term  arises in inhomogeneous magnetic field and
$\overline{d}$ -  is the vector-arm of the applied torque relative to the totation axis.

The torque $\textbf{L}$  leads to twisting of the elastic
thread until the forces of its elastic deformation do not
compensate  the applied torque. The angle of the thread rotation $\phi$  is detected, for example, by
a  capacitance changes.

For a small twisting angle $\phi \ll \theta_0 $, the restoring mechanical moment
of the twisted elastic thread is $L_\varphi= M B \sin\theta_0$, where $\theta_0$ is the angle
between the field direction  and the normal to the plane.
While the deviations from the equilibrium are small, $\varphi <10^{-4}$рад,
torsion scales operate linearly with  $\varphi \propto M$.

The authors \cite{eisenstein_APL_1985, eisenstein_PRL_1985}  estimated the sensitivity
limit for thread twisting as $1\mu$rad, and the magnetometer in total - as
$10^{-12}$J/T (or $10^{-9}$CGS) in field of 5\,Tesla,  that is equivalent to $10^{11}$  Bohr magnetons.
For detecting  the de Haas van Alphen effect (dHvA) with such a relatively  low magnetometer sensitivity
the authors used a GaAs/AlGaAs heterostructure containing
 a large number of parallel connected
2D electron sheets with a total area of 2\,cm$^2$ \cite{eisenstein_APL_1985}, and even
 12.5\,cm$^2$ \cite{eisenstein_PRL_1985}. Due to the nonlinearity of capacitance changes
with angle (disc misalignment), the
the amplitude $\textbf{m}$  was measured in
Ref.~\cite{eisenstein_APL_1985, eisenstein_PRL_1985} with
25\% uncertainty.

A different design of the torsion magnetometer with a capacitor more sensitive to the angle of sample  rotation
 was developed by Templeton \cite{templeton_JAP_1988} and was later  applied with some improvements
 in a number of works \cite{jones_SSC_1995, matthews_RSI_2004, matthews_PRB_2004}. In this design,
 twisting the thread with the sample and the capacitor plate cause changes in the effective capacitor gap $d$ rather than the plates area. As a result,  relative capacitance changes  amounts  $\delta d/d$, rather than $\delta S/S$, as in the design Fig.~\ref{fig:eisenstein-torsion}, giving a gain in the threshold
sensitivity by an order of magnitude.

Due to the small gap $d\approx 0.2$mm between the capacitor plates, the magnetometer threshold resolution,
in terms of the rotation angle $(1/C)(dC/d\theta)$,  in this design could be
a factor of  $\approx 25$ better than in the magnetometer Fig.~\ref{fig:eisenstein-torsion}, though in practice it appeared to be improved only by 10 times being limited by vibrations.
Another advantage of this design is the possibility of applying electrostatic (ponderomotor) force, by applying a DC voltage between the capacitor plates. Such a  feature is useful for damping the rotation system dynamics,  for calibrating absolute value of the
elastic torque (in situ, in the course of experiment), as well as for introducing a  feedback and,
thereby linearizing the amplitude response characteristics of the magnetometer.

Threshold resolution was $10^{-12}$J in terms of the detectable torque,
 and  $\sim  1\mu$rad in units of
the  detectable rotation angle, or $10^{-5}$
- relative change in capacitance. This resolution enabled
detecting dHvA oscillations for a single
heterojunction of  8\,mm$^2$ area,  with a total number
of electrons $\sim 7\times 10^{10}$.

Since all torsion magnetometers are based on a freely suspended electromechanical system,
the main source of noise are vibrations. Wiegers et al. \cite{wiegers_RSI_1998, wiegers_PRL_1997}
described the design more resistant to vibrations because
it contains a cylindrically symmetric rotor capacitor,
and the sample is located in the center mass of the rotary unit.

These design features have reduced a parasitic link coupling with external vibrations. The
resonant frequency of the suspended system is 1.5Hz, and the sensitivity threshold of
this magnetometer can be estimated from the reported measurements of the
oscillatory signal as $\delta m \sim 0.01\mu_B$  per electron,
although slow variations of the background were a factor of
10  larger \cite{wiegers_RSI_1998, wiegers_PRL_1997}. The authors
estimated the threshold magnetometer sensitivity as
$10^{-13}$J/Tesla, which is equivalent to $\delta M = 10^{10}$ Bohr magneton
 in field of 1\,T.

\subsubsection{Torsion magnetometers with optical detection
}

In torsion magnetometers with capacitive sensors, the detecting bridge circuit is fed
with a  low frequency AC  voltage that
can induce unwanted emf at the sample contacts.

In Refs.~\cite{schaapman_APL_2002, bominaar_NJP_2006}, an optical
 technique was used for sample deviations detecting. For this purpose,
a laser beam was introduced into a cryostat via a multimode fiber,
  reflected from the sample and then reached photodetector.
   The magnetometer was successfully used \cite{schaapman_PRB_2003} for measuring
  electron magnetization  of quasi-two-dimensional organic
small crystals (weighing 0.13 \,mg), as well as
for magnetization measurements with GaAs double quantum
wells \cite{bominaar_NJP_2006}, and  single layer
GaAs/AlGaAs  heterostructure \cite{schaapman_APL_2002}.
Threshold sensitivity
$2\times 10^{-13}$J/Tesla in field of 15\,T  corresponds to
magnetization changes  $\delta M = 5\times 10^{-3}\mu_B$ per electron.
Optical
detection turned out to be workable even for
measurements in the field of Bitter magnet, which
creates a fairly large electrical noise. In this case, the threshold sensitivity
was albeit lower by an order of magnitude, but still was enough for
studying quantum oscillations of magnetization
single-  \cite{schaapman_PRB_2003} and double-layer heterojunctions
GaAs/AlGaAs \cite{bominaar_NJP_2006}.

\subsubsection{Microconsole-type magnetometers}
The operation principle of  these magnetometers is similar to
the torsion balance. Just like in the latter, the
torque acting on a sample from the magnetic field, is balanced by mechanical
torque of elastic forces. The difference is that
the elastic element undergoes bending deformation rather than
torsion.

In Ref.~\cite{whiskermagnetometer-PTE_78}, a ``flexural''
magnetometer is described, in which the sample is not integrated
in a single process with the console, but itself plays a role of the  bending element.
Thus, the torque acting on the sample causes
 bending the sample itself, and not the auxiliary elastic
element.
The sample -  a flat threadlike crystal (whisker)
$\sim 1\mu$m thick and $l\sim 200 - 1000\mu$m long
is placed in magnetic field tilted by $0<\varphi<90^\circ$ relative its plane.

One end of the sample is firmly fixed. Magnetic field induces the torque  $\textbf{L}= \nabla_\varphi (\textbf{MB})$. According to the elasticity theory,
the resulting bending of the sample, characterized by
 a certain average angle $\alpha$ is equal to:
\beq
\alpha= \kappa \frac{Ll}{Ed^3b},
\eeq
where $E$ -the elastic modulus, $d$ is the thickness, $b$-sample width,
and $\kappa$ - is a factor of the order of unity.

As can be seen from this equation, the bending angle
is inversely  proportional to  $d^3$, while the torque $L$ itself
is proportional to the sample volume, i.e. thickness $d$.
Thus, with a decrease in the sample thickness, the bending angle does not
decreases, but increases! The measured bending angle is
is a measure of the mechanical torque $L$, and therefore
of the magnetic moment {\bf {M}}.

Using this magnetometer, quantum
magnetization oscillations were measured for a threadlike
Bi crystal with sizes $1\times 10\times 600\,\mu$m$^3$ \cite{whisker-jdPhys_78}.
The bending angle for such
sample in this case was $10^{-3}$rad, and
magnetic moment changes $10^{-11}$\,CGS.

In order to  linearize the characteristics in the device,
a  feedback is introduced by applying voltage $U$
between the sample and a closely located metal
plane. The torque of electrostatic forces acting on the sample $\eta U^2$ compensates for
the measured torque  $L$. For a large amplification factor $\eta$
in the feedback circuit, the angle practically does not
vary, and
the $\eta U^2$ value is the measure of the sought for magnetization signal $dM/d\varphi$.
Due to the feedback, the dynamic range
the measured moment was 4 orders of magnitude,
i.e. 80\,dB \cite{whisker-jdPhys_78}. The noise level of the magnetometer
was $\sim (10^{-6}-10^{-7})$\,dyne$\cdot$cm in the 1\,Hz bandwidth.

\begin{figure}[h]
\begin{center}
\includegraphics[width=170pt]{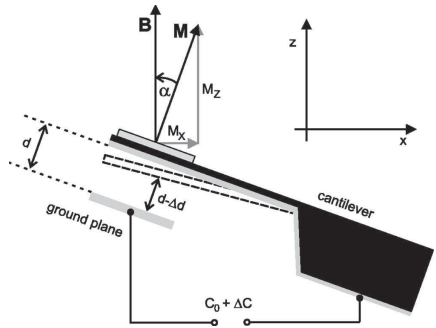}
\caption{Schematic design  of the  microconsole magnetometer. After Ref.~\cite{wilde_PRB_2005}}
\label{fig:fig1-wilde_PRB_2005}
\end{center}
\end{figure}

With the development of microtechnologies at the end of the past century,
 micromechanical cantilever (or console)
magnetometers (MCM) have been designed, based on both silicon - \cite{naughton_PhysicaB_1998,wilde_PRB_2005},
and GaAs technology \cite{harris_APL_1999,schwarz_APL_2000}. Figure~\ref{fig:fig1-wilde_PRB_2005})
shows a flat sample with 2D
electron gas mounted at the end of the elastic
microconsole. External magnetic field $\textbf{B}$  is applied
at an angle to the  sample plane.

Since   orbital electron  magnetization vector  is perpendicular to
the 2D electron gas plane,
console with the sample experiences  a mechanic torque
$\textbf{L}=\textbf{M}\times \textbf{B}$.
Thus, the sought for  magnetic moment, in the first
approximation, is proportional to the angle of deformation of
the elastic beam.

The operation principle  of the micro-console magnetometer
is illustrated in Fig.~\ref{fig:fig1-wilde_PRB_2005} from \cite{wilde_PRB_2005}:
the sample, glued at the end of the console, is a substrate of  1mm$^2$ area
 with Si/SiGe heterojunction containing 2D
electron system; for reducing weight, the
SiGe substrate is thinned to $10\,\mu$m.
The console with the sample is directed at an angle $\alpha$ relative to the magnetic field  $\textbf{B}$ vector;
 changes of the magnetic moment, proportional to
the torque $L$, are directly related with the
console bending angle: $\delta M= \delta L/B_\perp\tan \alpha$.
Typical thickness of the bending element - beam - is $10\mu$m \cite{schwarz_APL_2000}.

The micro-console magnetometers provide a  high sensitivity. In particular, in Ref.~\cite{schwarz_APL_2000}
the threshold sensitivity was $\delta M \approx 3\times 10^{-15} $J/T,
($\delta L\approx 1\times 10^{-14}$Nm) or
 $\sim 10^7 \mu_B$, i.e. $10^{-3} \mu_B$  per electron. In experiment \cite{wilde_PRB_2005}, using this
magnetometer, quantum magnetization oscillations  for 2D electron system in SiGe were reliably detected
 starting from field of 1\,T.

Due to this, in Ref.~\cite{wilde_PRB_2005} the authors were able to study  Landau levels  broadening, valley
and spin splitting and their renormalization  in magnetic field. In experiments  \cite{schwarz_PRB_2002,wilde_PRB_2006} with
 GaAs/AlGaAs heterojunction, the authors measured the density of
states profile  at the Landau levels — minima of the density
 of states, amplitude of oscillations in absolute
units, as well as enhancement of the spin splitting
caused by exchange interaction between the Landau levels.
In the experiment with
ZnSe/Zn$_{1-x-y}$Cd$_x$Mn$_y$Se quantum wells \cite{knobel_PRB_2002}  the evolution of
extended states at the Landau levels with levels broadening was studied.

In most of the magnetometers  \cite{naughton_PhysicaB_1998,wilde_PRB_2005,schwarz_APL_2000,schwarz_PRB_2002,wilde_PRB_2005},
bending of the micro-console  was detected via changes in the capacitance of the capacitor,
 $C=C_0+\delta C$  with a gap between the plates from $100\,\mu$m \cite{wilde_PRB_2005},  50\,$\mu$m
\cite{schwarz_PRB_2002} to $0.1\,\mu$m \cite{harris_PRL_2001}.

Beside the capacitive method of measuring
beam deformation, in a number of MCM designs  \cite{harris_PRL_2001, ruhe_PRB_2006}
an optical technique was used, similar to that considered
above for the  torsion magnetometers. Exclusion of electrical measurements of capacitance at AC
current allows, in principle, to get rid of cross
interference on the sample and, due to this, enables simultaneously
with magnetization measurement to measure DC transport-
 properties.

For measurements of the magnetic moment for a ferromagnet
semiconductor Ga$_{1-x}$Mn$_x$As  in  \cite{harris_APL_1999}] was used
another way -- to measure the shift of the vibration eigen  frequencies
of the console. The elastic beam in the magnetometer  \cite{harris_APL_1999} had a
transverse dimensions of $50\times 0.1\,\mu$m$^2$,
$400],\mu$m length, resonant frequency 1600\,Hz and the Q-factor
11500. A sample with dimension $40\times100\mu$м$^2$ was installed at the end of the beam.
With these parameters, a threshold sensitivity was shown  to
be  $3\times 10^6\mu_B$ in field of 0.1\,T and in the 1Hz bandwidth \cite{harris_APL_1999}.

The obvious advantage of MCM is their miniature design,
as well as a short response time, since
 the resonant frequency of the beam is approximately inversely
proportional to its length. In practical constructions,
the resonant frequency is $\sim 1$\,kHz \cite{naughton_PhysicaB_1998},
 allowing use of such magnetometers for measurements in
pulsed fields. Of course, for applications even in
 ``long'' pulse magnetic fields, of the order
of tens of ms, the mass of the sample should be small, of the order
of mg; generally, for static measurements, the mass is limited
by $\sim 10\,$mg, due to  unbalanced gravity. In some of these devices \cite{harris_APL_1999,schwarz_APL_2000, ruhe_PRB_2006},
the micro-console is integrated into a single unit with the sample
 -- GaAs-based heterojunction substrate with
two-dimensional electron gas.

It is worthy of  noting a related to MCM
local magnetometry technique -- scanning
magnetic force microscope (MFM), to be briefly considered
further in section \ref{MFM}.

\subsubsection{Vibrating-type magnetometer}

In vibrating magnetometers (VM), the measured signal is the
EMF induced by mechanical vibration of the sample relative to a pick-up coil,
placed in a constant magnetic field. Vibrating
magnetometer was invented by S. Foner \cite{foner_1956} and
described  by him in detail in Ref.~\cite{foner_1959}. In the original design
\cite{foner_1956, foner_1959} the sample vibration was driven
by  the laud speaker cone in the direction perpendicular to magnetic field. Threshold
 sensitivity in terms of the susceptibility the author estimated
 as $\delta\chi/\chi \sim 2\times 10^{-10}$ in the frequency bandwidth $2\times 10^{-2}$\,Hz
 \cite{foner_1959} and in terms of  magnetization -- as $10^{-9}$CGS in field of 1\,Tesla \cite{foner_1996}.

Review \cite{foner_1996} considers various options of the pick-up systems
for detecting  the induced AC magnetic field, including SQUID magnetometers.
There are also considered various examples of  VM in cryostats
with  3He pumping, dilution refrigerators,
and in hydrostatic pressure cells \cite{guertin_PRL_1976, guertin_LT_1978}. An
``inverse'' design of a VM is described in
\cite{reeves_1972}, in which the sample is likewise placed in the bore between
two coils. However, the coils are used not for receiving the induced voltage, but generate
an alternating magnetic field. As a result, the sample experiences a
 force causing its vibrations which are detected with piezo-sensors.

When  superconducting coils are used (in
contrast to the electromagnet with a gap between the magnet poles
as was in the first works \cite{foner_1956, foner_1959, foner_1996}) in the modern vibration
magnetometers \cite{springfort_1971,johansson_1976, hoon_1988, ausserlechner_1996, nizhankovskii_2007}, the sample moves  parallel to the magnetic  field of the solenoid, rather than perpendicular. Figure~3
shows a schematic design of the pick-up coils system for this geometry called ``vibrating sample magnetometer'' (VSM).

 Let a sample with a magnetic moment $M$  be placed at an average distance $Z$ from the plane of
the pick-up coil with a radius $r$. The sample sizes are presumed to be much less than $r$ and $Z$. The reciprocal
 motion of the sample along magnetic field $z(t)= Z_0\cos(\omega t)$ induces an emf in the pick-up
coil  \cite{springfort_1971, braggt_1976, johansson_1976, hoon_1988, ausserlechner_1996, nizhankovskii_2007}:
  $$
  E\propto \frac{\partial \Omega}{\partial Z}=
  6\pi r^2 Z\left(Z^2+r^2 \right)^{-5/2},
$$
  where $\Omega$-  is the angle, surrounding the coil perimeter from the point of sample location, and the
 transverse dimensions of the coils are assumed to be much less $r$ and $Z$. For a pair of identical opposite-connected coils,  spaced $2Z$  apart along the  solenoid axis,
 the sample deviation from the center by the distance $x$ induces an emf:
 \begin{eqnarray}
  E(Z,x) &  \propto  & 6\pi r^2 [(Z+x){(Z+x)^2+r^2}^{-5/2}\\
 &+ &(Z-x){(Z-x)^2+r^2}^{-5/2}]
\end{eqnarray}

A detailed analysis of the emf induced in the pick-up
coils  of various geometries and for their various location
is given in Ref.~\cite{ausserlechner_1996}. The amplitude of the induced voltage
has a maximum at $Z=r/2$, however for
achieving the most ``flat'' characteristics $E(x)$
(weakly sensitive to radial deviation  of the sample middle
point from the ideal position) usually
 $Z=\sqrt{3}r/2$  is chosen \cite{springfort_1971}.

For driving  the sample vibration, several techniques are now
 used: electric motors with lowering gears \cite{johansson_1976}, bimorph piezo-elements \cite{mangum_1970},
crankshaft mechanisms \cite{hoon_1988} and stepper motors \cite{nizhankovskii_2007}.
The threshold sensitivity of VSM typically ranges from $\sim 10^{-5}$CGS \cite{johansson_1976, nizhankovskii_2007}  to $10^{-6}$CGS \cite{mangum_1970}.

With an increase in the amplitude of oscillations, even harmonics appear in the
picked-up EMF in receiving coils. In Ref.~\cite{ausserlechner_1996}, the amplitude of the second harmonic
was used for absolute calibration
 of VSM; this method is conceptually similar to the one discussed below --
finding the  magnetization amplitudes for a nonlinear
oscillator \cite{pudalov74}.

\begin{figure}[h]
\begin{center}
\includegraphics[width=120pt]{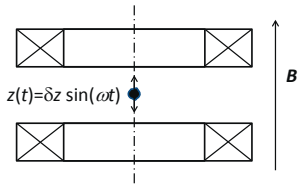}
\caption{Schematics of pickup coils  in VSM for sample vibrations  along the field}
\end{center}
\label{fig3}
\end{figure}

Vibrating sample magnetometers  have proven to be
 reliable and convenient tools and until now
are produced by a number of manufacturers
of scientific equipment \cite{cryogenic}.

\subsubsection{Summary}
The described above electromechanical and electromagnetic magnetometers
provide possibility of taking measurements of the thermodynamic magnetization $M=-dF/dB$ in absolute units. For the majority of them (except some console-type),
\cite{naughton_PhysicaB_1998})
magnetization measurements  are performed under slow magnetic field sweeping  or at a constant field.
 For lowering unwanted vibrations effect, all these instruments have a low resonant frequency of their elastic mechanical system,  $\sim 1$\,Hz, and therefore field modulation is not used.
As a result, the magnetometers are rather slow  and beside the oscillatory magnetization  detect an unwanted slow monotonic background signal, related with drifting environment parameters, drifting the construction and  magnetism of its elements,   etc.
Despite all these shortcomings, due to the simplicity of design, VSM are widely used and  are commercially available for laboratory applications.

Obviously, to reduce the effect of drift,
it is necessary to increase the modulation frequency of the signal. The only parameter
 allowing fast modulation is the concentration of electrons that can be
changed in 2D structures by varying the gate voltage.
This modulation method is described below in section
 \ref{gate_modulation}.

\subsection{Electromagnetic type magnetometers}

\subsubsection{SQUID-magnetometer}

The first measurements of the magnetization quantum oscillations  for
2D electron sheet were  done in Refs.~\cite{stormer-dHvA_JVST_1983, havasoja_SSci_1984}, namely using the SQUID-magnetometers. By now, a large number of laboratory instruments are described in the literature, besides, the SQUID-magnetometers are commercially available; for this reason we don't consider, but only briefly mention them here. In Refs.~\cite{stormer-dHvA_JVST_1983, havasoja_SSci_1984} the authors used a commerical SQUID-magnetometer, whose threshold sensitivity was insufficient for detecting quantum oscillations of a single two-dimensional layer of electrons with density
 $\sim 10^{11}$cm$^{-2}$.
For this reason, the authors used a set of  23 parallel  GaAs-AlGaAs heterostructures,
each with  173 two-dimensional layers  (quantum wells), in total
$\sim 4000$ of parallel connected 2D layers of the  240\,cm$^2$ area.
As a result, the authors for the first time observed quantum oscillations (de Haas-van Alphen effect)
for  2D electron system.

In a more advanced design \cite{meinel_APL_1997}
the threshold  sensitivity was improved more than
by three orders of magnitude; as a result, the authors registered
 quantum oscillatory magnetization
for electrons in a GaAs-AlGaAs heterostructure with 7\,mm$^2$ area.
Subsequently, they were able to study electron
magnetization in the fractional quantum
Hall effect  regime \cite{meinel_PRL_1999}.

In the SQUID magnetometer \cite{meinel_APL_1997}, a thin
film SQUID sensor was used with integrated multi-turn
superconducting coil. The first-order gradiometer
was connected to the input superconducting coil, creating a flux transformer.
Sample with 2D
electron system was positioned in one of the reception
pick-up loops of the gradiometer. SQUID itself was located in
a remote cryostat  and was shielded
from the stray magnetic field. To reduce noise, measurements
 were carried out by modulating the gate voltage  of the 2D heterostructure,
 at a frequency of 1.2\,kHz, at
which the SQUID noise level was the smallest.
In the absence of a field, the noise level of the magnetometer was
$ 3.5\times 10^{-5}\Phi_0/\sqrt{\textrm{Hz}}$, (for $\approx 2\times 10^{10}$  electrons
in the sample), however, the noise increased with field, by approximately a factor of
10  already in the field of 6\,Tesla.

\subsubsection{Modulation technique of magnetic susceptibility measurements}
\label{gate_modulation}

The total orbital magnetization for a typical
sample of a $10^{-2}$cm$^2$   area with the number of electrons $2.4\times 10^9$
in the field  $B =10$\,Tesla amounts to  only $\sim 4\times 10^{-11}$CGS.
In order to measure such a small quantity it is necessary to use the above rather complex
electromechanical constructions, poorly compatible
with  magnetic field modulation. For solving this technical problem, Fang and Styles
\cite{fang-stiles_PRB_1983}
 modulated  the electron concentration, rather than external magnetic field.
Implementation of  the much higher frequency
reduces the most difficult problem of a
low frequency $1/f$ noise. In the experiment \cite{fang-stiles_PRB_1983}
the gate voltage  of  the gated structure was modulated at
a frequency of 100\,kHz, and for receiving
 the induced signal, a thin film
coil was fabricated on the surface of the insulating Al$_2$O$_3$ layer, deposited atop
the Al gate.

In order the picked-up alternating  magnetic field would not be shielded by the conducting polysilicon gate,
the latter was  lithographically split into  20 strips, each $25\,\mu$m wide. Under harmonic modulation of the gate voltage
 $V_g=V_{g0}+\Delta V_g \cos(\omega t)$, and, accordingly, modulation of
electron concentration in the two-dimensional layer $\Delta n_s=(1/e)C\Delta V_g\cos(\omega t)$, the
 voltage $V(t)$  induced in the receiving coil
 is proportional to the oscillatory component $dM/dn_s(B, n_s)$:
 \beq
 V(t) = \frac{d\Phi}{dt}=\frac{SC}{e}\left|\frac{dM}{dn_s}\right|\frac{dV_g}{dt}=
 S\left(N+\frac{1}{2}\right)\hbar\omega C\Delta V_g/m^*c,
 \eeq
where $\Phi$ is the magnetic flux across the pick-up coil, $S$ -- total area of the two-dimensional channel,
   $C$ -- capacitance of the gated MOS structure,
   $m^*$-  electron effective mass, and $N$ is the Landau level number.

The latter equation has no fitting parameters;
sample dimensions, area and capacitance are easily determined. Despite this apparent simplicity,
measurement of the absolute amplitude of oscillations with this technique is impeded
by the  recharging time of the MOS structure
$\sim C/\sigma_{xx} \sim C\rho_{xy}^2/\rho_{xx}$ which for correct amplitude measurements
should be much smaller than
the modulation period. In practice, this requirement can
hardly be fulfilled, especially when approaching the
quantum Hall effect where the conductivity
of the 2D system drops exponentially (and therefore,
recharging time increases)
in a two-dimensional structure \cite{pudalov_SSC_1984}).

\subsubsection{Oscillatory magnetization measurements in a system with nonlinear magnetization}
In Ref.~\cite{pudalov74}, a method was proposed and implemented
for measuring the amplitude of the  electron
magnetization oscillations  from quantum oscillations of any
other  quantity (specifically, for example, magnetostriction),
under nonlinear  conditions of magnetic interaction.
 The parameter measured in this method is the shape
or the spectrum of quantum oscillations; it does not decrease
proportionally to  the sample volume, this feature in principle
does make it applicable to systems with a small number of
electrons. The method is based on the fact that
despite the smallness of the magnetization oscillations $\delta M$,
i.e. the amplitude of the dHvA effect, the oscillation period
for large Fermi surfaces is also small
$\delta B/B\ll 1$ and therefore the differential magnetic
susceptibility $|\partial M/\partial B|\sim \delta M/\delta B$  becomes comparable
 with $1/4\pi$. As the result, magnetic induction
in the sample $B$ differs significantly from the external
field,  $B= H+4\pi(1-D)M$.
This difference causes the so called ``magnetic interaction'' or ``Shoenberg effect''
 \cite{shoe_68} (here, $D$   is the demagnetizing factor). Magnetization $M$ is determined
self-consistently by solving the exact nonlinear equation  \cite{pippard}
\beq
M \sim \sum_{r=1}^\infty A_r \sin\left( \frac{\omega}{H+4\pi(1-D)M}\right),
\label{Eq:pippard}
\eeq
where $\omega=cS/e\hbar$ - is the circular oscillation frequency for the given extremal FS cross-section $S$,
 and $r$ is the oscillatory harmonic number in the Lifshitz-Kosevich formula \cite{LK_JETP_1955,
isihara-JPC_1986}.
In Ref.~\cite{pudalov74} this equation was solved by successive  approximations and
the amplitude of magnetization quantum oscillations (orbital electron magnetization)
has been determined  by comparing spectrum of the measured oscillations with
solution of  equation (\ref{Eq:pippard}).

The described above pioneer experiments \cite{stormer-dHvA_JVST_1983, fang-stiles_PRB_1983, pudalov74}
have demonstrated a possibility of measuring orbital electron magnetization in non-magnetic metals and semiconductors, however, in view of the complexity of the methods used,
and their inherent shortcomings,  in the future they were little  used.

\section{Electron spin susceptibility
from charge transport measurements}
\subsection{Spin susceptibility
from monotonic magnetotransport in the in-plane field}

In order to get information on the  spin susceptibility
of electron systems from monotonic magnetotransport,
measurements are  performed (a)  in strong
fields ($g^*\mu_B B_\parallel \sim E_F\gg  k_BT$), or (b)  in weak fields
($g^*\mu_B B_\parallel \ll k_BT$).\\

\subsubsection{High field measurements}

The first method is based on the empirical fact that for the  ideal (zero thickness) 2D system, the in-plane  magnetic field    couples
only with the spin degree of freedom. When magnetic field reaches the  complete spin polarization value,
 the magnetoresistance
of a 2D system exhibits a feature  (in Si-MOS and
Si/SiGe — the magnetoresistance saturates) \cite{gold-dolgopolov_JETPL_2000, shashkin_PRL_2001, pudalov_PRL_2002, tutuc_PRL_2002, gold_PRB_2003, das_PRB_2005, gao_PRB_2006, lu_PRB_2008};  from the position of this feature
in a number of works, the renormalized
 spin susceptibility  $\chi^*\propto g^*m^*$ value was determined. Here $g^*$ and $m^*$ are the renormalized g-factor and  effective mass of electrons, respectively.

\begin{figure}[h]
\begin{center}
\includegraphics[width=220pt]{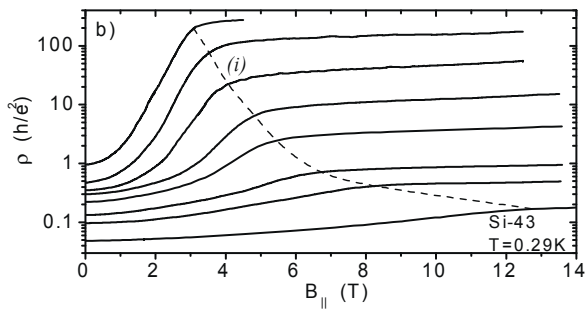}
\caption{$\rho_{xx}$  as a function of magnetic field $B_\parallel$  for 2D system of  electron in silicon at $T=0.29$\,K. Dashed line marks the field of magnetoresistance saturation. Density value  (from top to bottom) $n=1.91$, 1.98, 2.07,
2.16, 2.25, 2.48, 2.70, 3.61 in units of $10^{11}$cm$^{-2}$. From Ref.~\cite{pudalov_PRL_2002}.
}
\label{Fig:R(Bpar)}
\end{center}
\end{figure}

The advantage of this method is the simplicity of
measurements and apparent simplicity of data interpretation. The disadvantages
are connected, firstly, with perturbative action
of strong fields which ``cut off''  temperature
dependence of $\chi^*$  \cite{altshuler_JETPL_1982}, secondly, with the field influence on the
 $g^*=g^*(B)$ value due to the nonlinear character of magnetization \cite{zhang-das_PRL_2006}, and thirdly, with
disorder effect on the measured magnetoresistance saturation field
\cite{pudalov_PRL_2002, vitkalov_PRB_2002, sara_JPSJ_2003}.
Nevertheless, several works  \cite{tutuc_PRL_2002} reported the consistency of the  $\chi^*$ values,  obtained from the spin polarization field and by other techniques, considered below.

\subsubsection{Measurements in low and zero field}
\label{e-e-corrections}

Quantum corrections to magnetoconductivity in weak $B_\parallel$ field originate from the dependence of the effective number of triplet channels of electron-electron interaction on Zeeman splitting.
From the magnetoconductivity  measured in weak field $\delta\sigma_{xx}(B_\parallel)$, or from temperature dependence of the conductivity in zero field $\sigma_{xx}(T)$ one can extract the quantum interaction corrections
\cite{altshuler_JETPL_1982}. According to theory  \cite{ZNA_PRB_2001}, their magnitude  depends on  $g^*=g_b/(1+F_0^\sigma)$  via the Fermi-liquid coupling constant $F_0^\sigma$ in the $e - h$  channel:
\begin{equation}
\label{corection_structure}
\Delta\sigma_{ee}(T, B=0)  =\delta\sigma_C(T) + N_t \delta\sigma_t(T, F_0^\sigma).
\end{equation}

In equation (\ref{corection_structure}), the first and second terms  describe e-e interaction correction  in the singlet and  triplet channels,
respectively, $N_t$ is the number of triplet channels ($N_t=15$ for (001)-Si in weak field  and for
 not too low temperatures  \cite{vitkalov_PRB_2003, klimov_PRB_2008}).  In the  ballistic interaction regime (of ``high'' temperatures) $k_BT\tau/\hbar \gg 1$
\begin{eqnarray}
\delta\sigma_C(T, B=0) & \approx (k_BT\tau)/\pi\hbar \\
\delta\sigma_t(T, B=0) &\approx \frac{k_BT\tau}{\pi\hbar}\frac{F_0^\sigma}{1+F_0^\sigma}.
\end{eqnarray}

For the low-temperature diffusive regime of interactions, $k_BT\tau/\hbar \ll 1$
the interaction correction depends logarithmically on temperature \cite{altshuler_JETPL_1982, ZNA_PRB_2001}.

Extracting quantum correction from transport in zero field is relatively easy performed
in the ballistic regime, from the measured quasi-linear $T$-dependence
$\Delta\sigma_{ee}(B=0) \propto T\tau$; as a result of such approach, a number of works \cite{shashkin_PRB_2002, coleridge_PRB_2002, kvon_PRB_2002, proskuryakov_PRL_2002, safonov_JPSJ_2003, noh_PRB_2003, noh_JPSJ_2003,  kvon_PRB_2003,  pudalov_PRL_2003, savchenko_Phys E_2004} reported measurements of the interaction-renormalized  $g$- factor as a function of the carrier density.

 For nonzero $B_\parallel$ magnetic field, and within the same ballistic regime, interaction quantum correction to magnetoconductivity
\begin{equation}
\Delta\sigma_{ee}(T, B_\parallel) \approx \Delta\sigma_{ee}(T) +4\Delta\sigma^Z(E_Z,T, F_0^\sigma),
\end{equation}
$$
\Delta\sigma_{ee}(T, B_\parallel)\propto \frac{1}{\pi}\left(\frac{2F_0^\sigma}{1+F_0^\sigma}\right) \left( \frac{g\mu_B B}{k_B T}\right)^2\frac{k_BT\tau}{\hbar}
$$
depends quadratically on field and inversely on temperature \cite{ZNA_PRB_2001, vitkalov_PRB_2003}, that in principle enables to determine  $g^*$.

However, the $g^*$-factor values determined in such way   from magnetotransport, as a rule, lead to  $F_0^\sigma$-values   not fully consistent with the ones, determined from $\sigma(T, B=0)$ dependence \cite{pudalov_PRL_2003, vitkalov_PRB_2002, klimov_PRB_2008}.
One of the reason is the dependence of theoretical expression for quantum correction on the character of disorder potential \cite{ZNA_PRB_2001, gornyi_PRB_2004, gornyi_PRL_2003}, which for the real  2D systems is poorly known
\cite{clarke_NatPhys_2007}.
Another cause of the discrepancies is related to the difficult disentangling  of the interaction quantum corrections from classical and  semiclassical magnetoresistance effects \cite{proskuryakov_PRL_2003, morgun_PRB_2016}.

In the   ``low-temperature'' diffusive interaction regime  $k_BT\tau \ll \hbar$,  in weak fields  $g/\mu_B B\ll k_B T$, according to theory  \cite{ZNA_PRB_2001}, quantum correction to the nagnetoconductivity is proportional to $1/T^2$:
$$
\Delta\sigma_{xx} \propto \left( \frac{g\mu_B B}{k_B T}\right)^2.
$$
Their disentangling from  the semiclassical magnetoresistance
represents rather hard  task \cite{kuntsevich_PRB_2013, morgun_PRB_2016}
(for more detailed discussing this issue - see \cite{kuntsevich_PRB_2013, morgun_PRB_2016, pudalov_JSNM_2017}).

The disadvantages of all considered in this section transport-type methods of $g^*$-factor
 measurements   is their indirect character; clearly,  their results depend on the
theoretical models, on simplifying assumptions, etc. Additional complicating factor is
the dependence of the spin
polarization field on disorder \cite{pudalov_PRL_2002, vitkalov_PRB_2002,  sara_JPSJ_2003}. Finally,
all the above methods enable to determine only
the renormalized $g$-factor $g^*$, whereas the effective mass $m^*$, needed to determine $\chi^*\propto g^*m^*$,
must be found from other measurements, for example
from temperature dependence of the  quantum
oscillations amplitude; the oscillatory  methods and effects are
discussed below.

\subsection{Spin susceptibility
from quantum oscillations in tilted magnetic field}

\begin{figure}[h]
\begin{center}
\includegraphics[width=210pt]{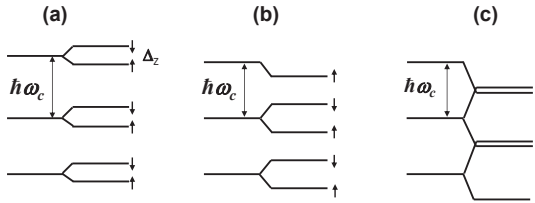}
\caption{Energy spectrum evolution for a single-valley 2D electron system
with Zeeman splitting  $\Delta_Z$: (a)  $\Delta_Z \ll \hbar\omega_c$.
(b) $\Delta_Z = \hbar\omega_c/2$,
(c) $\Delta_Z \approx \hbar\omega_c$. Vertical arrows depict electron spin polarization.}
\label{fig:spectrum}
\end{center}
\end{figure}

The simplest  and most widely used method of the spin susceptibility measurements for two-dimensional electron systems \cite{fang_PRB_1968, okamoto_PRL_1999, zhu_PRL_2003, shayegan_PRB_2008} was suggested and first implemented by  F. Fang and P.J. Styles \cite{fang_PRB_1968}. It consists of
magnetoresistance oscillation measurements (SdH effect) in magnetic field,
tilted from the direction normal  to the  2D system plane.
The method is based on the fact, that the cyclotron energy $\hbar\omega_c$
 is related only with magnetic field component   $B_{\perp}$, perpendicular to the 2D system plane.
In its turn,  Zeeman splitting of the Landau levels $\Delta_Z=g \mu_B B$ depends on the total magnetic field
$B_{\rm tot}$. In semiconductors, the  $g$-factor value  is often close to 2, and effective cyclotron mass $m^*\ll m_e$ is small; therefore, the Zeeman energy in purely perpendicular field is usually
small as compared with the cyclotron gap:
$\hbar \omega_c/\Delta_Z= (2/g)(m_e/m^*) (B_\perp/B_{\rm tot})$;
this case is schematically shown in Fig.~\ref{fig:spectrum}\,a.

As magnetic field is tilted, the perpendicular field component, that enters the orbital effects, decreases,
whereas Zeeman splitting remains constant.
At a certain tilt angle  $\theta$, Zeeman splitting becomes equal  to the half of the
cyclotron splitting (see Fig.~\ref{fig:spectrum}b),
and the observed oscillation frequency doubles. This condition (so called ``spin-zero'')
enables to determine  the spin susceptibility value for the known tilt angle $\theta$
as  $\chi^*/\chi_b=\cos\theta/0.38$
\cite{okamoto_PRL_1999}. Here
$\chi^*/\chi_b=g^*m^*/g_b m_b$, $\chi^*$, $g^*$, and $m^*$  are the interaction-renormalized spin susceptibility,
$g$-factor Lande and  effective mass, respectively;
 $\chi_b$, $g_b=2$  and
$m_b=0.19m_e$ --  are their band values for (100) Si.
This method is applicable in case of the spectrum
shown in Fig.~\ref{fig:spectrum}\,a,b  and  cannot be applied for
$\chi^*/\chi_b > 1/0.38=2.63$,  i.e., for larger Zeeman splitting, as shown on
Fig.~\ref{fig:spectrum}\,c.

\subsection{Spin susceptibility
from quantum oscillations interference in vector field}
\label{crossedfield}

An alternative, more flexible  technique for quantum oscillations measurements in magnetic field
with electrically controlled magnetic field vector was implemented in Ref.~\cite{cross_field}.
Nowadays,  vector magnets are commercially available and are not rare.
Magnetic field component in the plane of the 2D system $B_{\parallel}$ produces unequal spin subband population, which is needed to determine  spin susceptibility value.
The normal magnetic field component is required for observing quantum oscillations related with   Landau level quantization,  
and, hence, for counting  electron population in each spin subband.

As mentioned above, the conventional way of oscillation measurements in tilted field \cite{fang_PRB_1968, okamoto_PRL_1999, zhu_PRL_2003} fails when Zeeman energy
exceeds half of the cyclotron energy and further field tilting cannot decrease the Zeeman
contribution \cite{gm_PRL_2002}.
The ``crossed-field'' measurement technique  with independently variable  magnetic field components is free of these limitations and
enables to expand the measurements range to the low density values $n_s$,
where Zeeman energy strongly increases due to spin susceptibility renormalization, as shown in  Fig.~\ref{fig:spectrum}\,c.

\begin{figure}[h]
\begin{center}
\includegraphics[width=115pt]{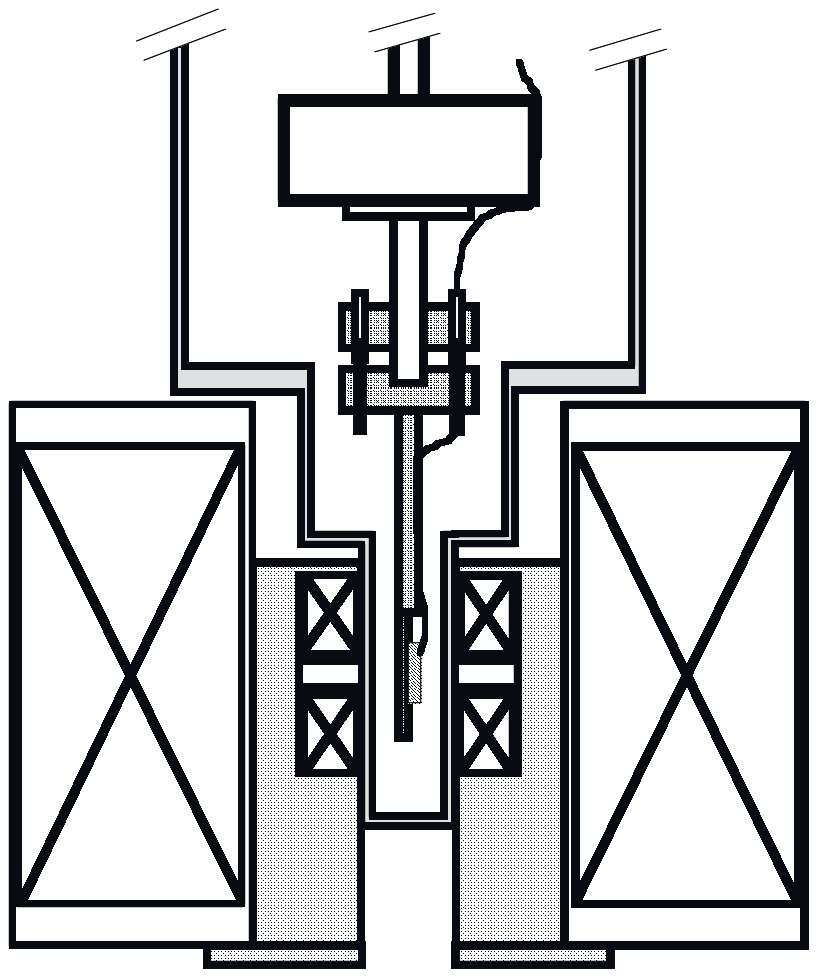}
\hspace{0.1in}
\includegraphics[width=110pt]{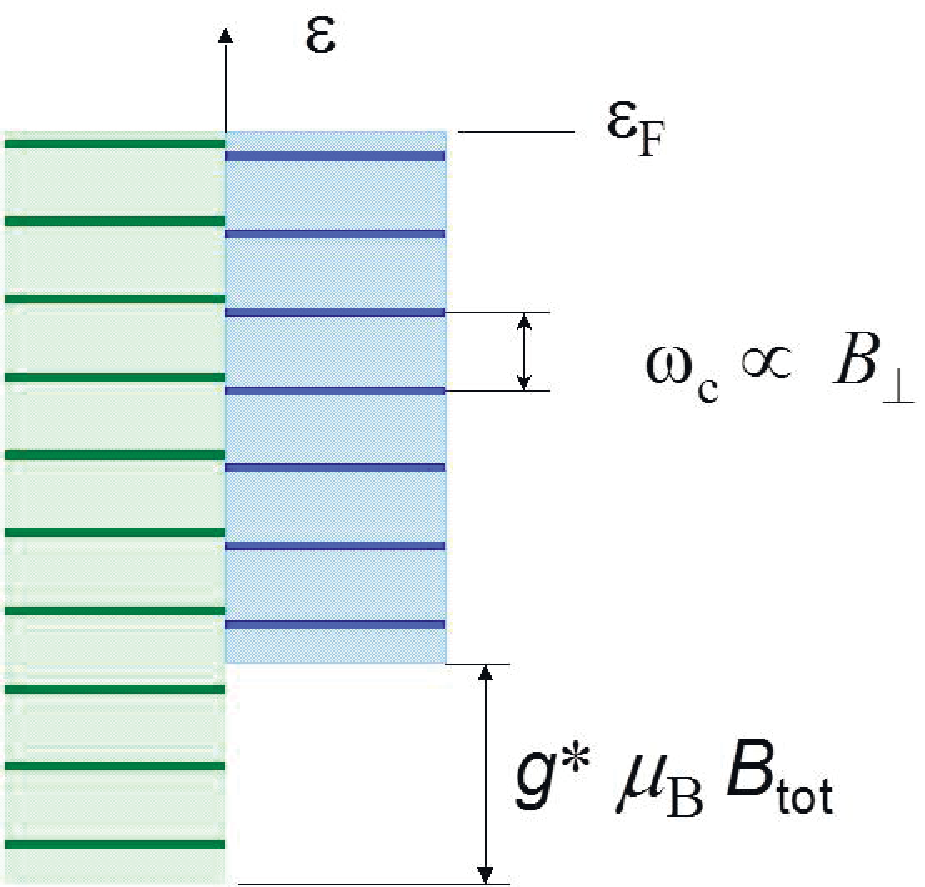}
\caption{Left -- the vector magnetic field setup with two crossed coils \cite{cross_field}.
The main superconducting magnet provides the in-plane magnetic field $B_\parallel$ up to 8\,T.
The second superconducting split coil system, positioned inside the main solenoid,
produces the  field  $B_\perp$ up to 1.5\,T normal to the 2D plane.
The sample  is  centered with respect to both solenoids and attached
to the cold finger of the mixing
chamber   $^3$He/$^4$He  \cite{cross_field, klimov_PRB_2008}.  Right -- schematic spectrum of the Landau levels
in two spin subbands, splitted by the  field $B_{\rm tot}$.}
\label{fig:cross-field}
\end{center}
\end{figure}

In the presence of a perpendicular field  $B_\perp$, the energy spectrum of a two-dimensional system is fully quantized and consists of equidistant Landau levels.
Application of the  $B_\parallel$ field induces beating of the quantum oscillations, which are registered as a
function of  $B_\perp$ field.
The cause of oscillation beating is explained on the right panel of Fig.~\ref{fig:cross-field}:
Zeeman splitting of the Landau levels induces nonequal population of the filled Landau levels in the
 $\uparrow$ and $\downarrow$ spin subbands.

The uppermost Landau levels in the two
spin subbands vary with the $B_\perp$  field at different rates. For some $B_\perp$ field values, they cross
the Fermi energy $E_F$   in phase and the oscillation amplitudes are summed up.
For other $B_\perp$ values  the Landau levels in two subbands  cross $E_F$  out of phase and the oscillation amplitudes  are subtracted. The beat frequency
is proportional to the spin polarization of the  2D
electron system \cite{gm_PRL_2002}:
\begin{equation}
P \equiv \frac{n_\uparrow - n_\downarrow}{n}
=\frac{\chi^* B_{\rm tot}}{g_b \mu_B},
\label{eq:polarization}
\end{equation}
where   $n_\uparrow, n_\downarrow$ stand for the populations of the  $\uparrow$ and $\downarrow$
spin subbands, respectively,
$g_b = 2$ is the bare  value of the Lande  $g$-factor for Si, and
$B_{\rm tot} = \sqrt{B_\perp^2 +B_\parallel^2}$.
For the degenerate  2D Fermi gas, equation (\ref{eq:polarization}) may be
written in  a way more convenient for practical use:
\begin{equation}
P= g^*m^* \frac{B_{\rm tot}}{\nu B_\perp},
\label{eq:polarization-1}
\end{equation}
where $\chi^*\propto g^* m^*$ --  is the Pauli
spin susceptibility of the Fermi liquid,
$g^*$, and $m^*$ are the renormalized
$g$-factor and effective mass, correspondingly, and
$\nu=nh/(eB_\perp)$ - Landau level filling factor.
One can see that the sought for spin polarization and spin susceptibility can be found from the beating period.

Of cause, for the interacting system, the shape and amplitude of oscillations  may differ from the simple Fermi liquid theory
\cite{LK_JETP_1955, isihara-JPC_1986, maslov_PRB_2003, adamov_PRB_2006},
specifically, for the strong inter-electron interaction, for strong overlapping and mixing of the Landau levels,
as well as  for breakup of the Fermi surface into the multi-phase state.
In particular, for the strong electron-electron interaction case,
the semiclassical Lifshitz-Kosevich
formula  \cite{LK_JETP_1955, isihara-JPC_1986}
is modified:
the interaction effects cause temperature- and magnetic field  dependent renormalization of $m^*$, and
$T_D$ \cite{adamov_PRB_2006, maslov_PRB_2003, pudalov_PRB_2014} in the exponential magnetooscillation damping factor.

These  complications, however, are insignificant for the beats analysis, provided that the parameters to be determined are only beating period and oscillation phase, i.e. spin polarization, and, in the end,  spin susceptibility.
Accordingly,  this technique enables to determine  spin susceptibility of {\it  delocalized electrons} possessing  sufficiently large relaxation time $\tau \gg 1/\omega_c$.

\begin{figure}[h]
\begin{center}
\includegraphics[width=160pt]{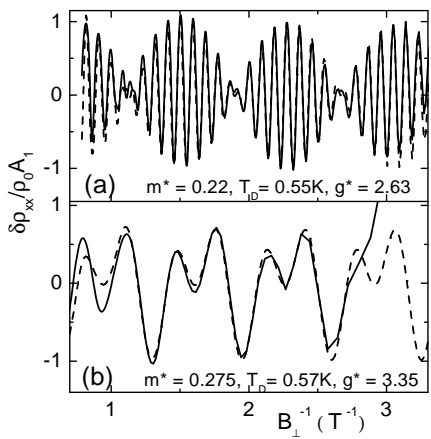}
\caption{Example of SdH  oscillation beating cited from \cite{pudalov_PRB_2014} for
 (a) $n = 3.76 \times 10^{11}$cm$^{-2}$, $T = 0.2$K, $B_\parallel = 2.15$T, $P =
20\%$; (b) $n = 1.815\times 10^{11}$cm$^{-2}$, $T = 0.2$K,
$B_\parallel  = 2.5$T, $P = 64\%$. The data are depicted with solid lines,
their approximation using equations \cite{LK_JETP_1955} (with the parameters shown)
- with  dashed lines. All data are normalized by the amplitude of the first  oscillation harmonic $A_1(B)$.}
\label{fig:SdH_beatings}
\end{center}
\end{figure}

\subsubsection{Comment}

All considered  in this section  techniques for  measuring  $\chi^*$
are based on  comparison of the  populations of the two spin subbands, i.e.  $M/B$. This certainly  differs from the true thermodynamically  defined quantity $\chi_T = dM/dB$, considered in the next section. In case when one and the same ensemble of electrons contributes to the measured quantity and when $M$ depends linearly on field,  $\chi^*$  and $\chi_T$ should coincide.
Besides, the measured susceptibility value is affected by the  non-ideality of the 2D system, such as finite thickness of the  2D layer  \cite{tutuc_PRB_2003, zhang_PRB_2005, palo_PRL_2005, zhang_PRL_2005} and magnetic field dependence of the susceptibility $\chi(B)$.

\section{Thermodynamic methods of measurements}
\subsection{Capacitive ``floating gate'' method   for chemical potential  measurements}

We consider here thermodynamic methods,
based on measurements of the chemical potential $\mu$ and
its derivative  $\partial\mu/\partial B$; these measurements are sensing practically overall
ensemble of charge carriers (including majority of the localized states),
capable of thermalizing within time interval of the order of seconds.
These methods are based on Maxwell relation for the second derivatives of the free energy $F$:
$$
\left( \frac{\partial^2 F}{\partial n\partial B}\right)
\equiv \left( \frac{\partial \mu}{\partial B}\right)=
-\left( \frac{\partial M}{\partial n}  \right)
$$

Method of measurements of the chemical  potential variations  $\delta \mu$ for 2D gated system
was put forward in Ref.~\cite{pudalov-quantosc_JETP_1985}; in fact, it is a version of the Kelvin technique.
 This method was used for measuring $\delta \mu$  as a function of magnetic field and electron density in a number of works
\cite{pudalov_valley-split_1985, pudalov-osc_EF_JETPL_1986, krav-NDOS_physlettA_1990, krav_evidence_PRB_1990}.

\subsection{Electrometric measurements of the chemical potential variations}
 \begin{figure}[h]
\begin{center}
\includegraphics[width=180pt]{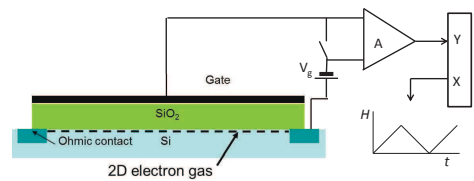}
\caption{Schematic setup for measurements of the chemical potential variations using  ``floating gate'' technique
\cite{pudalov_JETP_1985}. A is the electrometer.}
\label{fig:floating_gate}
\end{center}
\end{figure}

The principle of measurements is illustrated in Fig.~\ref{fig:floating_gate}. The 2D electron layer in MOS-structure is located near the Si surface and, together with the metallic gate forms a plane capacitor,
spacing between the electrodes is filled with silicon dioxide.
When a positive potential
$V_g$ is applied to the gate relative the 2D layer (via one of the ohmic contacts to the 2D layer) a charge is induced in the 2D layer, with a magnitude equal to the charge on the gate but of the opposite sign.

If the $V_g$ voltage source is disconnected from the gate, then at low temperatures leakage currents
are practically absent and the MOS structure keeps charge  $Q=C V_g$ for sufficiently long time. Hence, the density of electrons in the 2D layer,
 $n=Q/eS$, remains constant  ($S$  is the area of 2D layer,
 $e$ - the elementary charge); for the same reason  remains constant also the Fermi energy (counted from the lowest size quantization level) $E_F=2 \pi \hbar^2 n/m^* g_v g_s$. Here $g_s=2$, $g_v=2$ -  are the spin- and valley- degeneracy at the  (100)Si surface \cite{ando-review}.

When a magnetic field is applied perpendicular to the  2D plane, the energy of electrons in the two-dimensional system is fully quantized and in the absence of impurities and electron-electron interaction, the energy spectrum  consists of $\delta$-like discrete levels
\beq
E=\sum_Т\hbar\omega_c(N+ \frac{1}{2}) \pm \Delta_v \pm \Delta_Z,
\label{Landau_ladder}
\eeq
where $\Delta_v$, $\Delta_Z$ - are the valley- and Zeeman splitting in the spectrum
\cite{ando-review},
$N$ - Landau level index,
$\omega_c= eB/m^*c$ -  cyclotron frequency, $m^*$ -  the electron effective (band)
mass in the periodic lattice potential.

Corresponding to spectrum Eq.~(12), the Fermi level  $\varepsilon_F$ must have only quantized values.
Taking account of the Landau level spatial degeneracy
$n_H =\Phi/\Phi_0= eB/c\hbar$, the number of the filled levels $i$ for a given electron density
$n$ is determined by the condition
$i n_H \leq n < (i+1)n_H$
($\Phi$ is the magnetic flux per unit area,  $\Phi_0$ -- the magnetic flux quantum).
When magnetic field varies, the Fermi level changes in a step-like fashion, jumping from the $i$-th to the $(i+1)$-th level.
Importantly, the chemical potential changes at a constant electron density  $n$, since the gate voltage circuit is disconnected and the recharging current doesn't flow.
Such behavior of the chemical potential is considered conceptually in many textbooks  on the solid state physics
 (\cite{abrikosov_vvedenie, ashcroft-mermin, kittel}) and is a prime cause of the quantum oscillations of
  magnetization
(dHvA effect), conductivity (SdH effect) etc.

Variations in $\varepsilon_F(B)$ are equal to the chemical potential variations,
which are detected by the electrometer in the disconnected circuit shown on the diagram of Fig.~\ref{fig:floating_gate}.
In experiments \cite{pudalov-quantosc_JETP_1985}, the magnetic field was swept repeatedly in a sawtooth fashion, whereas the electrometer signal
was accumulated coherently  with the multichannel analyzer for  signal averaging in time domain.

For accurate electrometric  measurements  of the potential variations, the
gate potential should not change during the measurements time ($t \sim 10^4$\,s).
This sets rather strict though feasible requirements
for the leakage resistance in the measurement circuit (Fig.~\ref{fig:floating_gate}):
$R \gg   t/C \sim 10^{13}$\,Ohm, where $C
\approx 1$\,nF - capacitance of the gated  structure  \cite{pudalov_JETP_1985}.

\begin{figure}[h]
\begin{center}
\includegraphics[width=190pt]{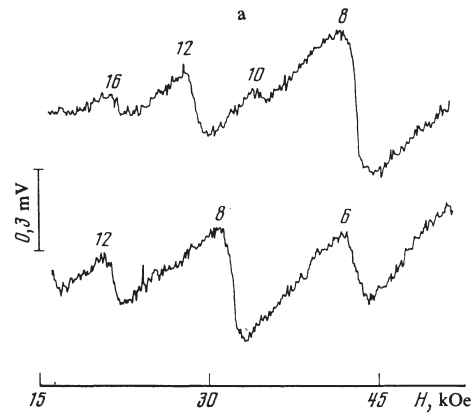}
\caption{An example of the measured chemical potential variations for
 two-dimensional layer of electrons in silicon MOS structure
versus magnetic field, from Ref.~\cite{pudalov_JETP_1985}.
The electron density is $8\times 10^{11}$\,cm$^{-2}$.
The bracket on the left side depicts magnitude of the effect.}
\label{fig:mu(H)}
\end{center}
\end{figure}

In order to implement the ``floating gate'' method, on the studied surface
 a  capacitive structure must be fabricated
  with a ``reference''  electrode, relative to which the chemical potential variations are to be measured.
  In Refs.~\cite{pudalov_JETP_1985, nizhan_PRB_1994, goldberg_PRB_1986}, the reference
electrode  was made of Al film (gate), deposited on top of the oxide, above the 2D layer. Typical oscillations of the chemical potential in magnetic field are shown in Fig.~\ref{fig:mu(H)}. The magnetic field derivative of the measured signal, evidently, equals to the changes in magnetization  per electron  $d\mu/dB =- dM/dn$.

The described above method was also used in Refs.~\cite{nizhan_JETP_1986, goldberg_PRB_1986}
for detecting chemical potential variations in the gated GaAs/AlGaAs heterostructure, and in Ref.~ \cite{semenchin_JETPL_1985, pudalov_valley-split_1985}
for measurements of  fine details of the electron spectrum in Si-MOS structure.
The attempts to measure with this technique chemical potential oscillations in
the bulk crystals of Bi \cite{nizhan_JETP_1986} and Be \cite{nizhan_JETP_1985} were unsuccessful,
possibly, because of the Fermi level pinning by the bulk carriers in three-dimensional crystal.
For thin YBCO-  and  Ni-films, variations of the chemical potential with magnetic  field
were successfully detected in Ref.~\cite{nizhan_PRB_1994}.

It is worth noting that this technique enables  probing properties of electrons of the near-surface layer
with thickness of the order of the Fermi wave length  (in case of the 2D layer of electrons
in a quantum well or in MOS structure), or of the order of the screening length - in bulk samples.
A modification of the ``floating gate'' technique with measurements of the DC-recharging current of the MOS structure was used in Ref.~\cite{pudalov-osc_EF_JETPL_1986} for measurements of the quantum oscillations of chemical potential as a function of the density of electrons in  2D layer.

\subsection{Modulation capacitive method of measuring chemical potential derivatives
}
\label{dmu/dB}

In the early 2000s,  the researchers interest shifted from orbital magnetization to the weaker spin magnetization effects, motivated by the issue of the potential Stoner instability in strongly correlated 2D electron system.
For spin magnetization measurements, in Ref.~\cite{prus_PRB_2003} an akin modulation method was developed   for thermodynamic magnetization  (MMTM) measurements, which was subsequently used in Refs.~\cite{shashkin_PRL_2006, anissimova_PRL_2006, teneh_PRL_2012}.

The measurements setup in  MMTM method is similar to that shown in Fig.~\ref{fig:floating_gate}. However, in order to exclude orbital effects in the spin magnetization measurements, magnetic field  $B_\parallel$ is applied parallel, rather than perpendicular to the 2D plane.
Modulation of the  magnetic field $B_\parallel$ at a low frequency
$\omega $ induces modulation of the chemical potential  of
2D electron layer  $\mu_{\rm 2D}$
and corresponding changes of the equilibrium charge.  In contrast to the diagram of  Fig.~\ref{fig:floating_gate}, here, a recharging current is measured in the capacitive structure.

\begin{figure}
\begin{center}
\includegraphics [width=180pt]{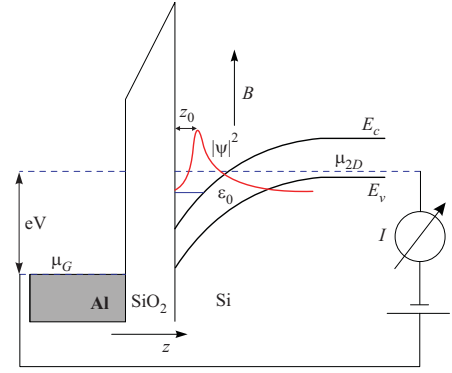}
\end{center}
\caption{Energy diagram and the principle of modulation-type measurements of spin magnetization
in magnetic field parallel to the 2D plane \cite{prus_PRB_2003, reznikov_JETPL_2010}.
$V$ is the voltage applied to the gate of MOS structure; $\mu_{2D}$, $\mu_{Al}$ are the chemical potentials of
 the 2D layer and the gate, respectively; $E_c$, $E_v$ - bottom of the conduction band and
ceiling of the valence band in bulk Si. $I$ is the current amplifier (recharging current is of the  fA-range).}
\label{dM-dB_setup}
\end{figure}

The principle of measurements is explained by Fig.~\ref{dM-dB_setup}.
The MOS structure is equivalent to a plane capacitor \cite{ando-review}.
Due to the overall sample electro-neutrality, the electron  layer charge
is exactly equal  (with an opposite sign) to the charge on the gate electrode.
When a DC voltage   $V$ is applied to the gate,
the free energy of the system becomes
\begin{equation} \label{VG}
F= F_{g} + F_{2D} -enV +\frac{e^2 n^2}{2C_0},
\end{equation}
where $F_{g}$, $F_{2D}$ are the free energies of the Al-gate film and the  2D layer, respectively.
The typical oxide thickness $d_{ox}\approx 200$nm,  whereas effective ``distance'' of the 2D layer from the interface is $z_0\approx 3.5$nm and remains almost constant, therefore  the capacitance $C_0$  in  Eq.~(\ref{VG}) differs only a little from the geometric capacitance of the classic capacitor,
$\sim (z_0/d_{ox}) \sim 1.7\%$.

\begin{equation} \label{dVG}
\frac{e^2}{C}\frac{d n}{d B}= -\frac{\partial \mu_{2D}}{\partial B} +
\frac{e^2n}{C^2}\frac{\partial C_0}{\partial B} \approx -\frac{\partial \mu}{\partial B}.
\end{equation}

The capacitor recharging current, $\delta I$ equals
\cite{prus_PRB_2003, reznikov_JETPL_2010, teneh_PRL_2012}:
\begin{equation}
\delta I=\frac{i\omega C_0\delta B}{e}\frac{\partial \mu
}{\partial B},
\end{equation}
where $\delta B$ is the amplitude of the magnetic field nodulation and
$C_0$ -- the capacitance
of the ``gate - 2D layer'' capacitor, measured independently by conventional capacitance bridge.
Contributions to the measured capacitance  due  to electron-electron interactions and finite width of the   2D
layer are negligibly small \cite{reznikov_JETPL_2010, teneh_PRL_2012}.

The quantity $\partial\mu/\partial B$ is found from the measured recharging current
and, due to the Maxwell relation $\partial M /\partial n=-\partial\mu/\partial B$, directly  renders the desired   ``magnetization per electron'' $\partial M/\partial n$.  The latter may be integrated with respect to $n$  to obtain the absolute value of the magnetization $M(B,n)$.
The magnetic susceptibility $\chi $ is calculated from the slope $M(B,n)$
as a function of   $B$ in low fields. A DC field, applied parallel to the modulation field
enables to
determine the nonlinear magnetic field dependence of $\partial M/\partial n$ and $M(n)$.

Importantly,  to the magnetization measured by this  method contribute all electrons,
capable of thermalizing  during the field modulation period  (of the 0.1 - 1\,s range) \cite{note1, gritsenko_UFN_2009}.
This difference in characteristic times (ps - in transport measurements and seconds
-- in thermodynamic measurements)
sets a fundamental difference in the character of information,
obtained from measurements with two different techniques. While in oscillatory transport measurements participate
only delocalized (mobile) electrons,
in thermodynamic measurements practically all electrons contribute, delocalized and localized.
The latter enables to carry thermodynamic measurements even in the insulator state, where the sample resistivity raises to the GOhm range.

In Ref.~\cite{reznikov_JETPL_2010} the applicability of this method
was justified for measurements
also in the regime of a complex capacitance, which acquires an imaginary part due to contact and channel
resistances; the latter enables to expand the range of applicability of the thermodynamic method
deep into the low density regime  of the insulator state.

Using  MMTM, in Refs.~\cite{prus_PRB_2003, shashkin_PRL_2006} magnetization per electron  $dM/dn$ was measured in high magnetic field for 2D  electron system in Si. As a result, features in magnetization anticipated at field of the full spin polarization were revealed.
Besides 2D electron system in Si, using this method, thermodynamic properties
of electrons were measured in   GaAs heterostructures \cite{tupikov_JETPL_2015, tupikov_NatCom_2015} and in HgTe quantum wells \cite{kuntsevich_JETPL_2020}.  The main physical results of these measurements are discussed in section  \ref{thermodynamic dM/dn}.

\section{Methods of local spin magnetization measurements}
Need in local methods of magnetic measurements emerged in relation with discovery of a whole class of akin  spin-orbit effects: the spin Hall effect (SHE), inverse spin Hall effect (ISHE), quantized spin Hall effect (QSHE), spin currents, etc. Studies in these directions  are related with development of spintronics,  particular,  semiconductor  spin logic elements, electric and optical means of the spin magnetization controlling  \cite{dyakonov_Spintronics_2008, sinova_RMP_2015, zvezdin_UFN_2018}, and, more generally -- with the need in effective information storage and computing devices.

The spin Hall effect reveals itself in accumulation of the spin polarization  at sample boundaries when electric current flows in the bulk; importantly, oppositely directed spins are accumulated at the opposite sample edges.
The idea of the spin Hall effect goes back to the anomalous Hall effect (AHE), which was observed already by E. Hall in ferromagnetic materials. In the absence of ferromagnetism, the spin-orbit interaction (SOI), the relativistic effect, also leads to the effects of spin accumulation, e.g., due to the  asymmetry of carrier deflection in the scattering processes \cite{barabanov_UFN_2015, zvezdin_UFN_2018}.
In ordinary Hall effect, Lorenz force deflects the charged carriers towards the sample edges, thus producing electric field directed perpendicular to the current. By contrast, in the anomalous Hall effect, SOI  produces the force, deflecting carriers to the opposite sample edges, depending on the spin direction.

\begin{figure}
\begin{center}
\includegraphics [width=190pt]{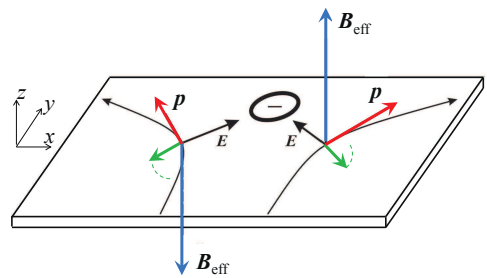}
\end{center}
\caption{Generation of the extrinsic SHE in a system with SOI.
Electrons moves along the  $xy$-plane  in a system with broken inversion symmetry
$z\rightarrow -z$  and are scattered by a negatively charged center.
 Red arrows show momentum direction, green arrows - equilibrium direction of spins in a system with Rashba-type spectrum.
 In the vicinity of the charged center  electrons are deflected by electric  field  $\textbf{E}$. In this process, the electron experiences  an effective  magnetic field ${\bf B}_{\rm eff} \propto [\textbf{p}\times \textbf{E}]$ (blue arrows), that is perpendicular to the
 $xy$ plane, and is inhomogeneous due to the momentum dependence. The gradient of the Zeeman energy  (of this effective field) forces spin rotation and their exit  out of the $xy$ plane as shown by the dashed arcs. The effective magnetic field is directed oppositely for electrons scattered to the left and right,  thus leading to accumulation of the spin magnetization with opposite directions  at the sample edges.
}
\label{fig:extrinsic SHE}
\end{figure}

The relationship between the charge and spin currents in non-ferromagnetic materials due to spin-orbit interaction was theoretically predicted in 1971 by M. Dyakonov and V. Perel \cite{dyakonov_PL_1971, dyakonov_ZhETF_1971}.
The idea of  experiment was suggested in Ref. \cite{averkiev_FTP_1983}, and the first measurements were done in \cite{bakun_JETPL_1984, tkachuk_JETPL_1986}.
This so called ``extrinsic''  SHE is related with an asymmetry of the electron scattering in the presence of   SOI and is an analogue of the  Mott scattering and deflection of electron beam in vacuum; its principle  is schematically  explained in Fig. \ref{fig:extrinsic SHE}.
The process of charge carrier scattering by  impurities includes a spin-dependent  difference of the deflection probability, that causes an unbalance  between oppositely directed spins.

\begin{figure}
\begin{center}
\includegraphics [width=180pt]{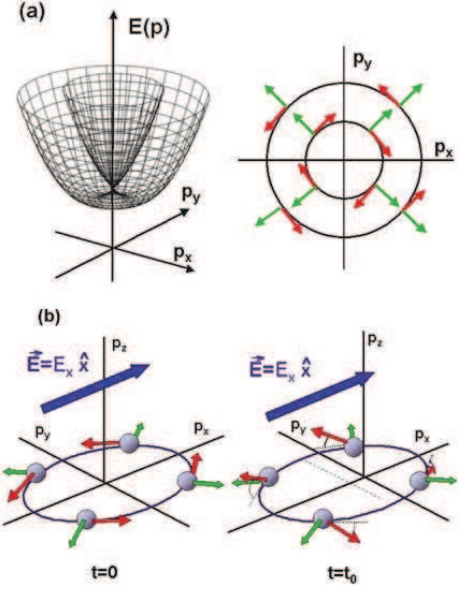}
\end{center}
\caption{(a) Energy spectrum  of electrons with  Rashba-type Hamiltonian for 2D system with spin-orbit interaction, left,  and the spectrum projection on the  $x-y$-plane  (Fermi surface, FS), right. Radially directed momenta are marked with green arrows on FS, the spin eigenvalues - with red arrows.
(b) Under application of an electric field in $x$-direction, FS  shifts by  $|eE_x t_0/\hbar|$  over the time $t_0 < \tau$ (where $\tau$ - the characteristic scattering time). When electron moves in momentum space in the presence of an electric field $E_x$, the effective torque
brings the spins  out of plane: upward for $p_y > 0$ and downward for $p_y < 0$,
thus causing the spin current in the $y$ direction. After Ref.~\cite{sinova PRL 2004}
}
\label{fig:intrinsic SHE}
\end{figure}

The ``extrinsic'' SHE was subsequently  supplemented  with the predicted \cite{hirsch_PRL_1999, zhang_PRL_2000}
strong  ``intrinsic''  SHE \cite{murakami_Science_2003, murakami_PRL_2004, sinova PRL 2004}, related with  dissipationless spin currents and irrelevant to electron scattering; its physical mechanism is illustrated   on
Fig.~\ref{fig:intrinsic SHE}.
The inverse SHE (ISHE), discovered in 2006
\cite{saitoh_APL_2006, valenzuela_Nat_2006, zhao_PRL_2006},
enables electric sensing of the spin current or spin magnetization gradient.
For experiments with  SHE, materials are selected  with high spin-orbit coupling parameters, such as  GaAs ($\lambda_s= 5.06$ e\AA$^2$),  ZnSe ($\lambda_s= 1.06$ e\AA$^2$), etc.

Several reviews are already published in this booming field,
including \cite{sinova_RMP_2015, ehlert_PSS_2014}; thanks to them, we avoid here  detailed consideration of the field, and only briefly describe the physical essence of the effects, experimental techniques and  the most remarkable results.

\subsection{Detecting local spin polarization}

\subsubsection{Detecting by optical techniques}
The problems in  SHE detection were initially caused by lacking of measurable electric signals;
for this reason the first experiments were done by optical methods
\cite{kato_Science_2004, wunderlich_PRL_2005, stern_PRL_2006}.  In experiments,  Kerr rotation of polarization was detected   (with spacial resolution) for the light transmitted through the epitaxial layers of
p-GaAs, n-InGaAs \cite{kato_Science_2004}, n-GaAs \cite{matsuzaka_PRB_2009, sih_PRL_2006},  n-ZnSe  \cite{stern_PRL_2006},  InGaN/GaN \cite{chang_PRL_2007} superlattices,  etc.

\begin{figure}
\begin{center}
\includegraphics [width=130pt]{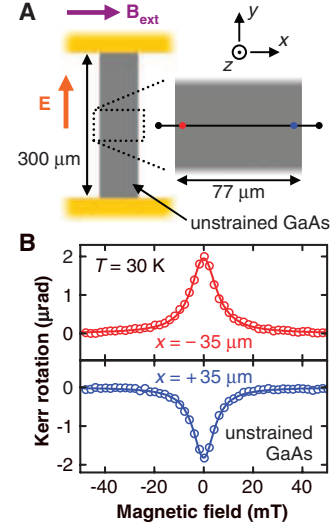}
\end{center}
\caption{The spin Hall effect in  unstrained GaAs, from Ref.~\cite{kato_Science_2004}. (A) Sample geometry.
(B) Typical measurement of the Kerr rotation versus magnetic field $B_{\rm ext}$
at two sample edges, $x= \pm 35\mu$m, for  $E= 10$mV/$\mu$m.
}
\label{fig:Kerr_detection}
\end{figure}

The polarization rotation indicates  electron spin accumulation at the sample edges, perpendicular to the  applied electric field.
Typical geometry of measurements is shown in Fig.~\ref{fig:Kerr_detection}.
The beam linear polarized  along $z$ was directed normally to the  plane of a rectangular sample,  and focuses into a spot about  $1\mu$m in diameter. The parameter to be analyzed is the polarization rotation angle of the reflected beam; it is proportional to the spin magnetization in  $z$ direction.  Such setup allows detecting angle-resolved  photoluminescence signal  at the opposite edges of the  2D hole system.  For precise sample positioning relative  the incident beam  in Ref.~\cite{matsuzaka_PRB_2009} a precise piezo-drive was used with 1$\mu$m coordinate resolution. In all measurements \cite{kato_Science_2004, wunderlich_PRL_2005, stern_PRL_2006, matsuzaka_PRB_2009} the Ti:sapphire laser with mode locking was used, with a typical  (0.15 - 1)\,ps pulse duration  and  76\,MHz repetition rate; the wavelength  825\,nm was tuned to the semiconductor absorption edge. In some experiments \cite{stern_PRL_2006} a pump-probe technique was used.

Results of the Kerr rotation measurements are shown in Fig.~\ref{fig:Kerr_detection}. The rotation angle corresponds to the  $z-$component of the spin polarization, which diminishes with the applied in-plane external magnetic field because of  spin precession.
The maximum Kerr angle is reached when the external field $B_{\rm ext}$ equals  the intrinsic spin magnetization $-B_{\rm int}$; this qualitative consideration helps to estimate the spin magnetization at edges. By taking similar measurements with uniaxially strained  InGaAs sample and и observing no Kerr rotation anisotropy, the authors concluded, that the observed effect in all cases was ``extrinsic'', rather than  ``intrinsic'' SHE. Analogous measurements were performed in \cite{kato_PRL_2004}  in the Voigt geometry with a beam transmitted through the strained epitaxial layers of InGaAs and GaAs. In all cases the authors observed similar magnitude of the rotation angle: $\sim 4\mu$rad for $E=4$mV/$\mu$m.

The above experiments were performed in the regime of a ``weak'' spin-orbit coupling, i.e. when the SO- splitting is smaller than the disorder-induced level broadening. In the ``strong''  SO-coupling regime  measurements were taken  in \cite{wunderlich_PRL_2005}, where the studied  2D hole layer was a part of a  {\em p-n}  junction in the light emitting diode. The current flowing through the {\em p-n}  junction is accompanied by electroluminescence due to electron-hole recombination. Beyond the ordinary exciton luminescence, the electroluminescence  spectrum  contained a circular polarized broadened line.  Because of the optical selection rules, the circular polarization in a certain direction  points at a spin polarization in this direction  of carriers involved in the recombination.

In all experiments, the Kerr angle magnitude $\theta_{\rm el}$ (or the spin accumulation magnitude $n_0$), in accord with the theory,   was found linear in electric field $E$. The spin relaxation time  $\tau$, extracted from data approximation  (e.g. in Fig.~\ref{fig:Kerr_detection}) was fit to a Lorentzian form $\theta_{\rm el} =\theta_0/[1+(\omega_L \tau)^2]$; it  didn't depend on $E$, but was coordinate-dependent, increasing with distance from edges.
At 20\,K the peak value of the spin density near the edges  was estimated as $n_0 \sim 16$\,spins/$\mu$m$^3$ \cite{stern_PRL_2006}.  Assuming a simple spin diffusion model, one can model the spin accumulation profile, related with spin current, as
$\theta_{\rm el}= n_0  {\textrm{sech}}(W/2L_s) \times \sinh(y/L_s)$,  where $L_s$ - is the spin diffusion length.
From approximation of experimental data in Ref.~\cite{stern_PRL_2006}, an estimate $L_s \approx 1.9\,\mu$m was found for $T = 20$K.
The spin current density along $y$  may be written as $|J_y^s| = L_s n_0/\tau$,  from which spin conductivity
$\sigma_{\rm SH}=-J_y^s/E_x \sim 3$(Ohm м$)^{-1}/|e|$.

 It is important to note  for potential applications, that while with temperature growth the magnitude of the effect diminished (as well as the spin polarization $n_0$,  spin relaxation time  $\tau$,  and spin diffusion length, the latter from $ 1.9\mu$m at 20\,K, to  $1.2\mu$m at 295\,K), the effect remained pronounced even at room temperature.

\subsubsection{Detecting by electrical methods
}

For electrical detection, set-ups with non-local geometry are used, in which spin-polarized carriers are injected from ferromagnetic to a nonmagnetic material.
The detecting method is commonly  based on the ISHE,  where the Hall voltage is induced by spin current.
Many experimental setups are described in literature \cite{valenzuela_Nature_2006, valenzuela_JAP_2007, saitoh_APL_2006, werake_PRL_2011, lou_NatPhys_2007, ehlert_PSS_2014, garlid_PRL_2010, brune_NatPhys_2010},
which use various nonmagnetic materials, including normal metals, superconductors, nanotubes etc.

Two different approaches are mainly used for nonlocal electric detecting of  SHE: (1) detecting the ``direct SHE'' i.e. spin accumulation at two edges of a sample due to SOI, under flow of charge current of unpolarized carriers, and detecting spin magnetization accumulated at the edges  with ferromagnetic potential contacts \cite{ehlert_PSS_2014, ehlert_PRB_2012, garlid_PRL_2010}, and (2) detecting  the ``inverse SHE'' (ISHE) by injection of polarized charge carriers via ferromagnetic current contacts and by detecting unbalance in spin accumulation at the edges with nonmagnetic potential contacts \cite{lou_NatPhys_2007, olejnik_PRL_2012, valenzuela_APL_2004, choi_NatNanotech_2015}.

\begin{figure}
\begin{center}
\includegraphics [width=260pt]{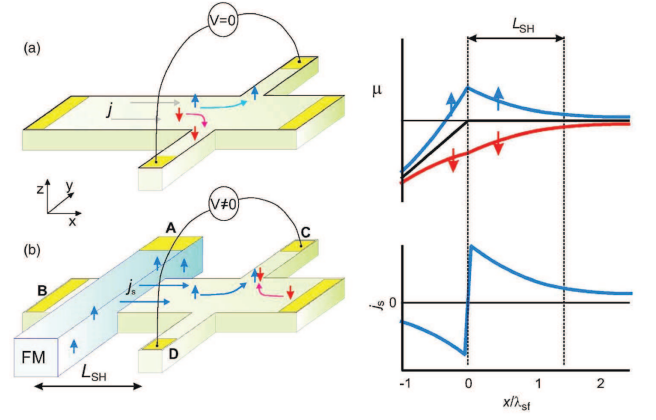}
\end{center}
\caption{
(a) Direct SHE: the spin unbalance arises at the sample edges due to SOI, when purely charge current flows. (b) ISHE: the Hall effect, induced by the spin current.
Purely spin current $J_s$ is injected from left to the right.
SOI causes separation of electrons with spins up and down, thereby inducing the transverse charge current and noticeable voltage.
Schematic coordinate dependences (c) of electrochemical potentials for spins $\uparrow$  and $\downarrow$,  when
the charge current flows from the ferromagnet to a nonmagnetic material from left to the right, and (d)  spin current $J_s$, related with  spin injection. Reproduced from Ref.~\cite{valenzuela_JAP_2007}.
}
\label{fig:schematics el-detection}
\end{figure}

Schematics of nonlocal electric detection by the 2nd method (ISHE) is shown in
Fig.~\ref{fig:schematics el-detection}).
In case the charge current is spin unpolarized (Fig.~\ref{fig:schematics el-detection},a), it generates spin accumulation at the sample edges (as well as in SHE), not leading to Hall voltage appearance because of the equal number of charge carriers deflects to opposite sides. However, in case the charge current is spin polarized (Fig.~\ref{fig:schematics el-detection},b) by means of ferromagnetic injection with  magnetization directed out of plane, the initial unbalance of electrons with spins $\uparrow$ and $\downarrow$  cause inequality of electrons, scattered to different sides. As a result, a Hall voltage arises between the Hall contacts  C and D. The Hall voltage is measured non-locally, away from  the injector, whereas Hall contacts and injector are disconnected galvanically in order to avoid voltages generated by ordinary Hall effect and by magnetoresistance anisotropy.
Therefore, the Hall effect, induced by the spin current, shown in Fig.~\ref{fig:schematics el-detection},b is the effect inverse to  SHE, shown in Fig.~\ref{fig:schematics el-detection},a.

\begin{figure}
\begin{center}
\includegraphics [width=90pt]{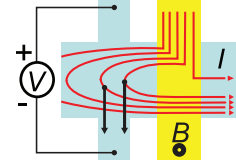}
\end{center}
\caption{Schematic current distribution in the vicinity of a ferromagnetic contact \cite{ehlert_PSS_2014}}
\label{fig:Fig5_Ehlert_review}
\end{figure}

The polarized electrons are injected in the vicinity of
$x=0$  and diffuse with equal probability  towards two opposite arms of nonmagnetic material. The process of nonlocal current flow is illustrated in Fig.~\ref{fig:Fig5_Ehlert_review}.
In the diffusion process, the nonlocal spin current  $J_s$ decays with distance away from the injection point as \cite{valenzuela_JAP_2007}
\beq
J_s(x)=\frac{P}{2}\left(\frac{I}{A_N}\right) \exp(-x/\lambda_{sf}),
\label{eq:spin-diffusion}
\eeq
where $P$ is the polarization of the injected current $I=I_{\rm AB}$ (Fig.~\ref{fig:schematics el-detection},b),
$A_N$ -- the cross-section area of the nonmagnetic strip, and $\lambda_{sf}$ is the spin diffusion length.
For the geometry shown in Fig.~\ref{fig:schematics el-detection},a
\beq
V_{\rm SH}=V_{\rm CD}= - E_y(x)/w_N=w_N \frac{\sigma_{\rm SH}}{\sigma_c^2} J_s(x),
\label{eq:Hall-voltage}
\eeq
where $w_N$- is the width of nonmagnetic metal strip, $\sigma_c$- Drude conductivity for the charge current and
$\sigma_{\rm SH}$ - the ``spin-Hall'' conductivity. Substituting Eq.~(\ref{eq:spin-diffusion}) to the Eq.~  (\ref{eq:Hall-voltage}), one obtains non-local Hall resistance $ R_{\rm SH}= R_{\rm AB,CD}=V_{\rm SH}/I$
$$
R_{\rm SH}= \frac{P}{2t_N}\frac{\sigma_{\rm SH}}{\sigma_c^2}e^{-x/\lambda_{sf}}
$$

\begin{figure}
\begin{center}
\includegraphics [width=160pt]{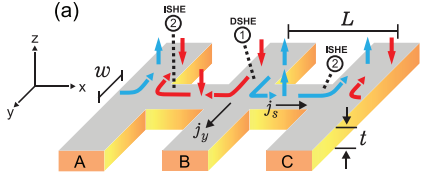}
\end{center}
\caption{Configuration of measurements with a double H- bridge, from Ref.~\cite{ehlert_PSS_2014}}
\label{fig:Fig6_Ehlert_pss_2014}
\end{figure}

In practical devices \cite{valenzuela_Nature_2006} CoFe was chosen as the ferromagnetic material, and Al - as a normal metal. The tunnel barrier between Al and CoFe is achieved by oxidation of the Al-strip.
The presence of the tunnel barrier is essential for uniform distribution of the injected current, and also for increasing  polarization of injected electrons.
Typical parameters of this device $P\approx 0.28$,
$\lambda_{sf} \sim 450$ and  $700$nm for the  Al- strip thickness of 12 and 25nm, respectively. The spin diffusion
length sets the  required strip length $L \sim 500-800$nm.

An elegant setup for the nonlocal ISHE detection  in the double-arm H-bridge was realized in Ref.~ \cite{brune_NatPhys_2010}. Usually, for ISHE detecting with two-arm bridges,  spin polarized carriers  are injected via a ferromagnetic contact \cite{choi_NatNanotech_2015}. Unlike this, for spin polarized current injection in Ref.~\cite{brune_NatPhys_2010}, a  HgTe/CdHgTe  heterostructure was used with HgTe quantum well thickness greater than 6.3\,nm. Due to the spectrum inversion, in such structure a regime of a topological insulator is formed, where the spin polarized current flows along the edges, thereby allowing to get rid of ferromagnetic contacts.

The idea of ISHE detection using the double H-bridge was suggested in Ref.~\cite{hankiewicz_PRB_2004}.  Such setup was used for measuring SHE  in  Au films \cite{mihajlovic_PRL_2009},  PbTe layers \cite{aleszkiewicz_PSS_2013},  and graphene \cite{balakrishnan_NatPhys_2013}. Principle of its operation is explained
in the inset to Fig.~\ref{fig:Fig6_Ehlert_pss_2014}.
A current of unpolarized charges $J_y$ flows in the middle arm  B. Under presence of SOI,
the dominant scattering direction depends on spin; as a result, a spin current $J_s$ arises in  perpendicular direction. Due to ISHE, carrier scattering causes charge current in the $y$-direction, perpendicular to the current $J_s$ (ISHE) and a difference of potentials (or current) is induced in arms  A  and  C. Despite the doubtless advantage of the double H-bridge, consisting in the absence of ferromagnetic contacts, the interpretation of results is hampered by the presence of side effects related with overheating of the arm B (due to Nernst–-Ettingshausen effect) and diffusive  transport \cite{ehlert_PSS_2014}.

\begin{figure}
\begin{center}
\includegraphics [width=170pt]{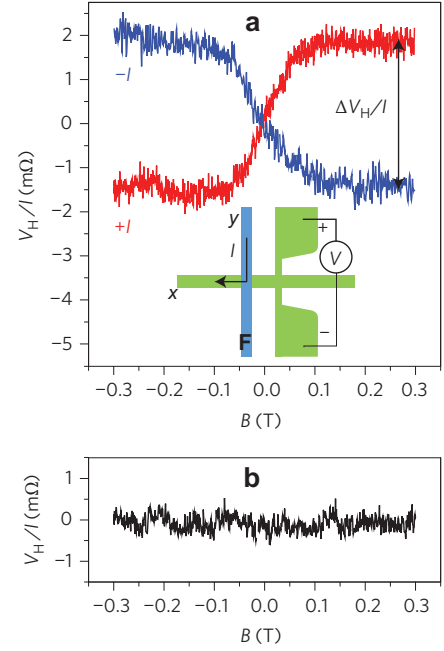}
\end{center}
\caption{(a) The ISHE signal versus external field $B$ applied  along  $x$, for the device with  InAs- channel $L=2.83\mu$m long, at $T = 1.8$\,К. F- ferromagnetic strip -- injector. The red curve: current $I>0$, blue curve: $I<0$. The inset shows geometry of measurements. (b) control measurement with $B$  applied along   $y$-axis, demonstrating the absence of the Hall voltage.
After Ref.~\cite{choi_NatNanotech_2015}}
\label{fig:Fig2_choi2015}
\end{figure}

The majority of the devices utilizes the extrinsic SHE, caused by the scattering anisotropy in the diffusive transport regime. The ballistic regime   of intrinsic SHE \cite{brune_NatPhys_2010, choi_NatNanotech_2015} was realized only for materials with large  carriers mean free path at low temperatures (e.g., InAs),  and for devices with a short  channel. Thus, in Ref.~\cite{choi_NatNanotech_2015} a method was utilized for spin precession detecting under ballistic propagation of carriers, injected from the ferromagnetic contact F  (Fig.~\ref{fig:Fig2_choi2015}) into the perpendicular strip of a nonmagnetic material with large SO coupling (InAs quantum well).

When carriers are injected from a contact polarized by the external field $B$ along $x$ direction, and are accelerated by an electric field to the left side into the  $x<0$ region, then in the nonlocal ballistic regime the charge current equals to zero in the region $x>0$ (see Fig.~\ref{fig:Fig5_Ehlert_review}).

In materials with Rashba spectrum, the spins tend to align perpendicular to the electron velocity $v_x$ and to the ``installed'' electric field $E_z$ of the quantum well. In this picture, the spin directed initially (at $x=0$)  along  $x$ starts precessing as a function of coordinate $x$. One can qualitatively think that both, the electron trajectory and Hall voltage $V_y\equiv V_H$  between the strip edges will exhibit spatial oscillations  \cite{datta_APL_1990}  with a period  $\lambda=\pi \hbar^2/\alpha m^*$,  where  $\alpha$ -  is the Rashba spectrum parameter.  As a result, the Hall voltage shows antinodes at distance $x=\lambda/4$, $3\lambda/4$,  etc. Its sign inversion under current inversion is seen in Fig.~\ref{fig:Fig2_choi2015}a.  For magnetization  directed along $y$, the carriers injected into InAs propagate ballistically with no spin precession, with no trajectory bending and the Hall voltage does not arise  (Fig.~\ref{fig:Fig2_choi2015},b).

Numerus experiments confirmed the operational capability of the described devices and the ability of electrical detecting SHE. Quantitative data were obtained  on the parameters of spin diffusion: the spin diffusion length $\lambda_{sf}$ and its temperature dependence
  \cite{valenzuela_JAP_2007, ehlert_PSS_2014}.

\subsection{Magnetometry based on NV-centers}
The methods described in the preceding section allow for sensing electron spin magnetization with spatial resolution  of the order of  $1\mu$m. However, in some cases there is a need in studying magnetization features on a nanoscale.
In magnetic materials and non-trivial magnetic phases, such as skyrmions, magnetic topological insulators, spin density waves, Abrikosov vortices in superconductors, the
non-uniform magnetic structures arise at  nanoscale. Negatively charged nitrogen-substituted vacancies (NV-centers) in diamond offer a possibility of sensing on the atomic scale, suitable for
 quantum magnetization probing with a nm-resolution.

\begin{figure}
\begin{center}
\includegraphics [width=250pt]{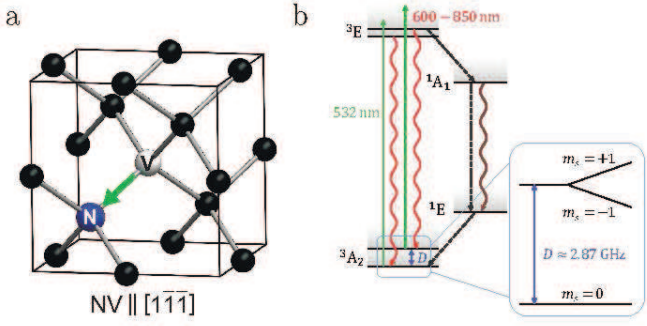}
\end{center}
\caption{(a) The diamond lattice, containing the NV-center, which is formed by the substituting nitrogen atom, neighboring to the carbon vacancy in the lattice.
The green arrow shows symmetry axis of the NV-center along [111] direction of the diamond lattice \cite{pham_NJP_2011}.
(b) Energy diagram of the  NV-center in diamond, containing zero field split electron levels with spin projection $m_s=0$ and degenerate levels $m_s=\pm 1$.
From Refs.~\cite{schloss_PRA_2018, barry_1903.08176}.
}
\label{fig:Fig1_barry_1903.08176}
\end{figure}

Figure \ref{fig:Fig1_barry_1903.08176},a  shows an  NV-center in the diamond lattice, and Fig.~\ref{fig:Fig1_barry_1903.08176},b -- schematic energy levels. The NV-center in diamond consists of the substituting nitrogen atom, neighboring to the carbon vacancy. Such centers emerge in bulk and nano-crystalline  diamonds:   synthetic diamonds grown by CVD, as a result of radiation damage and anneal, or by ion implantation and anneal. The centers exist as negative (NV$^-$) and neutral  (NV$^0$)  charge states \cite{doherty_PhysRep_2013}.
Besides diamond, the vacancy centers have been found also in  silicon carbide  (SiC) \cite{kraus_SciRep_2014, kraus_NatPhys_2014}.

Schematics of the  NV$^-$-center and energy level structure are shown in
Fig.~\ref{fig:Fig1_barry_1903.08176}. The ground ($^3A$) and excited ($^3E$) states form the triplet with sublevels $m_s=0$ and $m_s=\pm 1$. The transition $^3A_2\rightarrow$ $^{3}E$
may be excited in optical wavelength range 450-637nm, and the fluorescence of the transition
$^3E \rightarrow$ $^3A_2$  occurs in the wavelength range 637 to 800\,nm.

\begin{figure}
\begin{center}
\includegraphics [width=250pt]{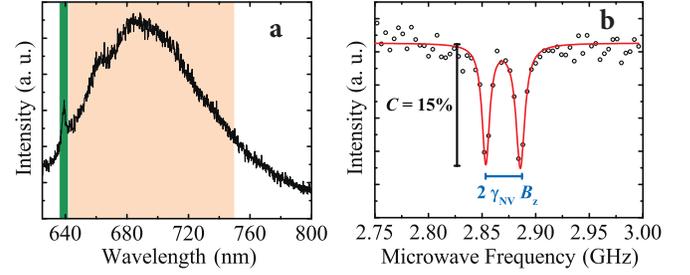}
\end{center}
\caption{(a) Typical luminescence spectrum  with a narrow ZPL and wide PSB.  (b) Example of results  of ODMR
shows two dips in the luminescence intensity,   located symmetrically relative to the zero field splitting of 2.87GCs. From Ref.~\cite{fuchs_NJP_2018}.
}
\label{fig:Fig1_Fuchs_NJP_2018}
\end{figure}

Figure  \ref{fig:Fig1_Fuchs_NJP_2018},a shows a luminescence spectrum at room temperature \cite{fuchs_NJP_2018}.
The purely electronic transitions between the excited $^3E$ and the ground $^3A$ states lead to a narrow zero-pnonon line  (ZPL) at  638\,nm. Beside this line,
there is a wide phonon wing  of lines (PSB), shifted to the red side; it contains about 96\% intensity of the  NV-center luminescence \cite{zhou_APL_2017}.

Optical transitions mainly occur with spin conservation, however, the spectrum contains levels crossings between the  singlet and triplet states. Therefore, beside the direct transition from $^3E$  to $^3A$,  the fluorescence decay channel includes also intermediate long-living singlet states, as well as  radiationless transitions from $^3E$ to $^1A_1$, and from $^1E$ to $^3A_2$. As a result, the relaxation rate to the  $m = 1$ state is higher, than to the  $m = 0$ state. Because of this difference,  under optical pumping an optical spin polarization develops - the main part of the population transfers to the  $m_s= 0$ state. The fluorescence of the NV-center is spin-dependent and its level is determined by the spin polarization degree. Such dynamics of the levels population
allows to polarize the electron spin of the  NV-center via a non-resonant excitation (typically, at the 532\,nm wavelength, by $1\mu$s pulses).

In the NV$^-$-center, the singlet and triplet spin sublevels  $m_s = 0$ and $m_s = \pm 1$ of the  $^3A_2$ ground state in zero field are split by the crystal field: the energy difference is $D=  2.87$GHz (Fig.~\ref{fig:Fig1_barry_1903.08176}).
Weak external magnetic field shifts the sublevels $m_s = \pm1$, so that their splitting varies
proportionally to the field projection  $B$ on the  NV-center axis:
$(1/hB)[E(m_s=\pm 1)-E(m_s=0)] =  ±2.8$MHz/Gs.
Therefore, the NV$^-$-centers may be detected not only in optical transition between the ground $^3A_2$ and excited  $^3E$ levels, but also in the microwave (MW) range,
using conventional electronic paramagnetic resonance (at a frequency of 2.87\,GHz in zero field),
or by optically detected magnetic resonance (ODMR) \cite{oort_SSP_1988, taylor_NatPhys_2008, maze_Nat_2008}.
In the latter case, the applied resonant microwave radiation transfers part of the population from $m_s= 0$, decreasing fluorescence signal, excited by non-resonant optical pumping.
The properties of NV-centers are reviewed in detail in Ref.~\cite{doherty_PhysRep_2013}.

Figure \ref{fig:Fig1_Fuchs_NJP_2018}  shows an example of optical detection  - optically detected magnetic resonance (ODMR) - a sharp intensity drop  of the narrow luminescence line under coincidence of the microwave signal frequency with spin subleveles splitting.
The unique distinction of the  NV- centers from other solid state systems with single spins
consists in that the long coherence time is achieved even at room temperature.
Thanks to this feature, the individual  NV-center in diamond crystal with low defect density may provide the threshold sensitivity as low as  30\,nT/Hz$^{1/2}$ \cite{maze_Nat_2008} and even
4.3\,nT/Hz$^{1/2}$ \cite{balasubramanian_NatMater_2009} at room temperature and in the atmospheric environment.

The second unique feature of the NV-center consists in the small  volume of the sensor, practically of an atomic scale.
This enables to bring the sensor to a sample at the nm-distance for visualizing magnetic field on nanoscale.
Magnetic field of individual spins decays as a 3rd power of distance and  were a sensor located away at  $\sim 1$\,mm distance, the field from the single spin would be
negligibly low,   $\sim 10^{-21}$\,Tesla.
The NV-centers may be formed within  5\,nm of the diamond surface,
conserving at the same time long enough spin relaxation time,
$100\mu$s \cite{ono_APL_2012}. The proximity of the NV-center to the diamond surface enables
sensing magnetic field of individual spins that is
in the range of $\mu$T \cite{maze_Nat_2008, balasubramanian_NatMater_2009, maletinsky_NatNanotech_2012, kleinlein_MicroelEng_2016, appel_RSI_2016}.

\subsubsection{Scanning microscopy based on NV-centers}
\begin{figure}
\begin{center}
\includegraphics [width=200pt]{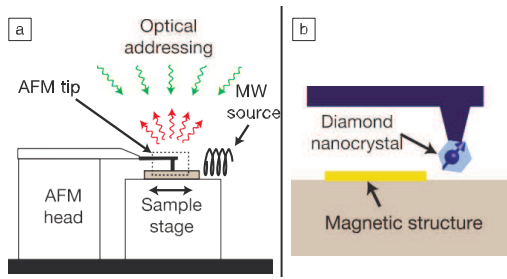}
\end{center}
\caption{(a) Schematics of the  NV-magnetometer \cite{maletinsky_NatNanotech_2012}.
(b) Schematics of the atomic force microscope (AFM) with a diamond nano-crystal probe,
containing a single NV-center
\cite{hong_MRSBull_2013}.
}
\label{fig:hong_MRSBull_2013}
\end{figure}

Magnetic sensors with NV-centers are compatible with the scanning probe microscopy technique; owing to this circumstance they are used for magnetic field visualization at nanoscales.
In the scanning magnetic  NV-microscope, the diamond nano-pillar serves not only as the probe tip, but also as a nano-photonic light guide. In the latter capacity it  effectively collects and guides photoluminescence signal from the NV-center to the optical  registration system \cite{rondin_RepProgrPhys_2014}. Schematic arrangement of the  NV-magnetometer with optical sensing is shown in Fig.~\ref{fig:hong_MRSBull_2013}. Theory of magnetic scanning  NV-magnetometers operation and  means of their  optimization  are considered in \cite{fuchs_NJP_2018, barry_RMP_2020}.

In this relatively young  area, several reviews and monographs are already published  \cite{jensen_book, doherty_PhysRep_2013, hong_MRSBull_2013, fuchs_NJP_2018, wojciechowski_Materials_2019, levine_Nanophotonics_2019, boretti_Beilstein J. Nanotechnol_2019, barry_1903.08176, schirhagl_ARPC_2014} and PhD-dissertations \cite{nizov_thesis}.
Particularly, there are described various applications of
the NV- magnetometry for studying ferromagnetic   50-nm grains
in magnetostatic bacteria, 10-nm grains in  meteorites,
magneto-marked cancer cells \cite{sage_Nat_2013, fu_Sci_2014, glenn_NatMeth_2015}. Owing to the high
 spatial resolution and  nontoxic diamond, NV-microscopes are successfully used in neuroscience and biology \cite{schirhagl_ARPC_2014}, including  intracellular dynamics detecting of a living cell   \cite{mcguinness_NatNano_2011, kuwahata_SciRep_2020}. The NV-magnetometers are expected to enable imaging of individual molecules by NMR and MRI techniques; detecting single electron spin was already demonstrated \cite{grinolds_NatPhys_2013}.

Traditional technical applications of the NV-magnetometers are the characterization of read/write magnetic heads, measurements of the stray fields from magnetic domains in  hard  disk drives, etc.
  \cite{maletinsky_NatNanotech_2012, jakobi_NatNanotech_2016}.
In condensed-matter physics, NV-magnetometry  was used for studying Meissner  effect,
structure of magnetic flux vortices in superconductors \cite{waxman_PRB_2014}, structure of domain walls and vortices in thin magnetic films  \cite{tetienne_Sci_2014, rondin_NatCom_2013}, spin-wave excitations \cite{sar_NatCom_2015}, skyrmions, spin ice, and other exotic materials \cite{rondin_RepProgrPhys_2014, dussaux_NatCom_2015}.

\subsection{Scanning probe magnetometers}
Beside the considered above magnetic microscopy based on  NV-centers,
more traditional methods are also widely used for local magnetic surface probing.

\subsubsection{Scanning magnetic force microscopes}
\label{MFM}

Since the first realization of the magnetic force microscope in  1987 \cite{martin_APL_1987}
a great number of  MFM designs have been developed and described in detail in literature \cite{yaminskii_UHN_1999, martin_APL_1987, cordova_NanoworldJ_2016}. To date, they have become common in laboratory practice and are commercially available as an option to the atomic force microscopes (AFM) \cite{ntmdt}.

The probes usually used for measurements with MFMs,  are  made of either magnetic materials, or with a magnetic film (Co) deposited  onto an ordinary  nonmagnetic probe \cite{Stiller_MeasSciTechnol_2017}. In the latter case, the stray magnetic fields in the vicinity of tip are smaller by an order of magnitude, than for the probes made of magnetic  wires.
In MFM measurements in static regime the probe  -- magnetic tip must be located away of the surface, in order the magnetic interaction forces exceeded Van der Waals forces (the dominant forces in the AFM regime). Because of this, MFMs have a limited spatial resolution. The MFM measurements usually require  two cycles of scanning: at a small distance of the surface and at a large distance, with subsequent subtraction of the results for excluding contribution from  Van der Waals interaction.

There are also developed bimodal MFM designs; they enable measuring the AFM and MFM signals during the single scan. For this purpose,  the small amplitude ($\sim 10$nm) mechanic oscillations are excited in the elastic console, simultaneously at two frequencies; by lock-in detecting the ac signals at two frequencies the two contributions are disentangled: from the the long-range magnetic forces (MFM), and short-range Van der Waals forces (AFM) \cite{jason_APL_2009, rodriguez_APL_2004, li_APL_2009, schwenk_APL_2014, schwenk_APL_2015}.

\subsubsection{Magneto-resonant force microscopy}
This method (MRFM) combines ESR and NMR methods with magnetic force microscopy \cite{sidles_APL_1991, sidles_RSI_1992}  and, in principle, allows for 3-dimensional imaging of magnetization inside materials; several reviews on MRFM are published, e.g., \cite{sidles_RMP_1995, yaminskii_UHN_1999}).
As well as MFM, MRFM contains an elastic console with a probe at its end, located at a small distance  from the sample. Microwave field  with a frequency tuned to the magnetic resonance changes spins orientation (of electrons or nuclear) and, hence, the sample magnetization. This causes changing the magnetic force acting on the sample and  shaking the  elastic console. In order to improve the MRFM sensitivity, the amplitude of the microwave field is modulated at the frequency of the console mechanical resonance; thereby the amplitude of its forced vibrations is the measure of the sought for magnetization.

When the probe is scanned relative the sample, the resonant vibration amplitude (of the Angstroem scale) of the cantilever  holding the sample is measured. This method is applicable for magnetic mapping  with  pumping either electron spins at the ESR frequency, or nuclear spins at the NMR frequency. In the earlier studies \cite{zuger_APL_1993}, a spatial resolution of  $\sim 5\,\mu$m was obtained. Later on, the spatial resolution was improved up to 0.9\,nm \cite{grob_Nanolet_2019}, whereas sensitivity -- up to 50-100 nuclear magnetons (for the $(3-5)$nm$^3$ voxel) \cite{rose_PRX_2018}.
Such magnetometers are now also commercially available \cite{zurich}.

\subsubsection{Scanning Hall microprobes}

Scanning Hall magnetometers have rather simple design, can operate in the wide temperature range and in atmospheric environment;  the commercially available instruments are  fabricated  by a number of manufacturers \cite{nano}.
As a Hall microprobe, semiconductor heterostructures are used with high mobility two-dimensional electron gas  in GaAs/AlGaAs \cite{chang_APL_1992}, InAlSb/InAsSb/InAlSb \cite{bando_JAP_2009},  as well as Bi \cite{sandhu_JJAP} and graphene \cite{sonusen_ASS_2014}.
For example, in Ref.~\cite{chang_APL_1992} a  Hall microscope is described with a field sensitivity $\sim 0.1$\,Gs and a spatial resolution of $\sim 0.35\,\mu$m, whereas  Ref.~\cite{dede_APL_2016} describes a vector magnetometer  with a $1\times 1\mu$m$^3$ GaAs-sensor, providing spatial resolution of  $\sim 700$\,nm.

\subsubsection{Scanning SQUID magnetometers}
The first scanning SQUID-magnetometer (SSM), or SQUID-microscope has been developed in  1992 \cite{black_APL_1993}.
 The operation principles of SQUID as the magnetic field sensor are described  in detail in the textbooks \cite{clarke-book, schmidt-book}. Typical SSM design includes scanning module  with a console,
which carries a micro-SQUID. In contrast with MFM, where the magnetic field spatial distribution is
deduced from the force acting between the probe and the sample, in SSM the magnetic field is measured with a superconducting pickup coil of the SQUID. Various designs of SSM are described in  review articles
\cite{kirtley_IBMJRD_1995, kirtley_RMS_1999}, and operation theory and data interpretation -- in \cite{reith_RSI_2197}.

For achieving high spatial resolution the most suitable are the direct current SQUIDs (dc SQUIDs).
Their pickup loop ($\sim 1-10\,\mu$m) and the SQUID sensor itself are fabricated using electron beam lithography
technique.
The threshold sensitivity is determined by the SQUID noise level and effective area of the pickup loop. For the
typical noise level $2\times 10^{-6}\Phi_0$/Hz$^{1/2}$  and the loop area  of 7$\,\mu$m$^2$, the noise level is   $10^{-6}$Gs/Hz$^{1/2}$.
In practical SSM devices \cite{vu_IEEE_1993, kirtley_APL_1995}, a spatial resolution of $\sim 20$\,nm and the lowest detected magnetic flux $(10^{-3} - 10^{-5})\Phi_0$/Hz$^{1/2}$ was achieved for the SQUID pickup loop diameter of  $\sim 1\,\mu$m.

The scanning SQUID microscopes are also available as commercial products  \cite{attocube}, in particular of the domestic design \cite{snigirev}.

\subsection{Comparison of the local magnetometry methods}
Each of the listed above local magnetometry methods has its own merits and demerits \cite{kirtley_RPP_2010}: MFM
possesses high spatial resolution (up to $10-100$nm) and can operate in a wide range of temperatures. The SQUID magnetometers have very high  sensitivity (up to $10^{-15}$T/Hz$^{1/2}$), but the worst  ($\sim 0.3-10\,\mu$m) spatial resolution and are capable of working only at low temperatures. The Hall microscopes have an intermediate resolution ($\sim 0.3-1\,\mu$m). The NV-magnetometers are characterized by a good combination of the spatial resolution ($\sim 1-10$nm), high magnetic sensitivity and a wide range of temperatures.
For all devices, however, there is a compromise between the accessible threshold sensitivity and spatial resolution: for example, for  NV-magnetometers the sensitivity raises sharply, up to pT/Hz$^{1/2}$ with NV-centers ensemble  (though with loss of the spatial resolution) in the $10^{-3}$mm$^3$ volume
\cite{taylor_NatPhys_2008, hong_MRSBull_2013, wolf_PRX_2015}.

\section{Results of the physical investigations}

In this section we briefly consider several key physical results,
obtained from measurements of the electron magnetization.

\subsection{Orbital magnetization of two-dimensional electron systems}
\label{osc.amplitude enhancement}
Beside the very fact of observing  dHvA oscillations in the 2D electron system, in experiments
\cite{stormer-dHvA_JVST_1983, havasoja_SSci_1984, eisenstein_PRL_1985, templeton_JAP_1988},
  the theory of magnetooscillations was tested for the two-dimensional  \cite{LK_JETP_1955,
isihara-JPC_1986, maslov_PRB_2003, adamov_PRB_2006} and quasi one-dimensional  \cite{wilde_PhysE_2004} systems.
In Refs.~\cite{meinel_PRL_1999}, electron magnetization was probed in the regime of the fractional quantum Hall effect.

Measurements of the orbital magnetization already in  1980-1990-th
were used for obtaining information on the disorder induced Landau level broadening,
their shape,  on the density of states within the gaps between the levels,  character of electron scattering \cite{potts_JPCM_1996, zhu_PRB_2003, usher09, ruhe_PRB_2006, wilde_PSS_2008},
on the spatial inhomogeneity of electron distribution and its effect on the oscillations damping \cite{shoenberg_book}.
Further, magnetization measurements in  2D systems were used for studying breakdown mechanisms
and current ``pinching''  in the regime of the quantum Hall effect (QHE).
The review article \cite{usher09} considers in detail the related contactless magnetic measurements for studies of the orbital magnetization and  charge transport.

Orbital magnetization measurements are also commonly used for  estimating
the residual resistance value in the quantum Hall effect regime.
The results of these measurements are briefly described in section \ref{hysteresis}.
Absolute amplitude of the dHvA oscillations, inter-electron exchange  interaction
at neighboring Landau levels were measured using various  methods of magnetometry \cite{pudalov-osc_EF_JETPL_1986, krav_PRB_1990, meinel_PRB_2001}. The results of these measurements were compared with
theoretical calculation \cite{macdonald_PRB_1986, meinel_PRB_2001} of the oscillations amplitude enhancement
due to the many-body effects of
electron-electron interaction (the so called ``inter Landau level interaction'').

Recently, in connection with intensive studies of quasi two-dimensional
high temperature superconductors and topological insulators, measurements
of the magnetization oscillations obtained even wider dissemination.

\subsubsection{Hysteresis non-stationary  recharging effects
 in the QHE regime
}
\label{hysteresis}
With growing  $\hbar\omega_c/k_BT$  ratio, the diagonal components of resistivity and conductivity in the QHE regime diminish exponentially  and further saturate.
The residual dissipative resistivity is the important parameter  both, for  clarifying transport mechanism
in the gapped state, and for estimation of the accuracy of reproducing the quantized Hall resistance in the Ohm standards \cite{pudalov_JETP_1985, krasnopolin_IET_1987}.
The residual resistance, though, is so tiny, that can hardly be measured
with contact-type transport techniques; besides, the
area in the vicinity of the heavily doped  contacts introduces an excessive electron scattering.
For this reason the possibility of  contactless estimation
of the true residual resistance using magnetometry is very valuable.

\begin{figure}
\begin{center}
\includegraphics [width=160pt]{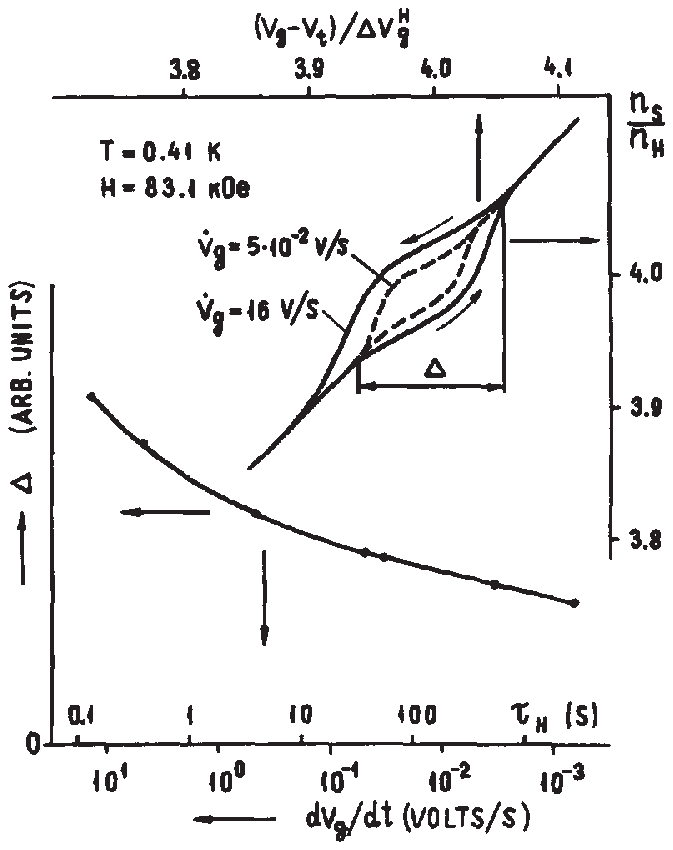}
\includegraphics [width=130pt]{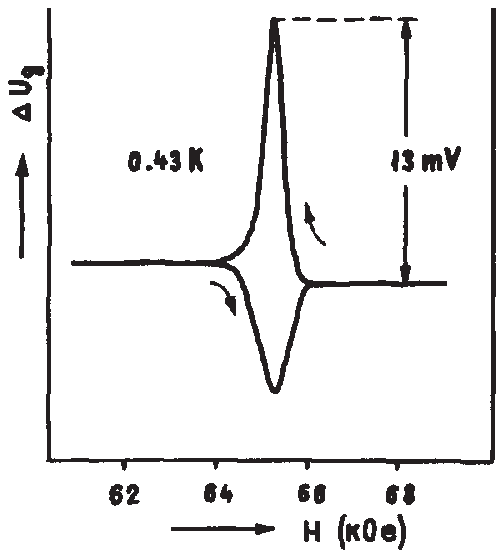}
\end{center}
\caption{Hysteresis variation of the chemical potential with
changes of the electron density in MOS structure
 (upper panel),
and magnetic field  (lower panel), cited from Ref.~\cite{pudalov_SSC_1984}. On the top right side the hysteresis loop is shown  for two  $dV_g/dt$ values.
On the bottom left  - dependence of the hysteresis loop width on the rate  $dV_g/dt$.}
\label{fig:pudalov-hysteresis}
\end{figure}

The nonstationary effects in recharging of the two-dimensional layer in the quantum Hall effect have been found
in Ref.~\cite{pudalov_SSC_1984} in measurements of the chemical potential  $\mu$ variations
for Si MOS structure, and, independently, in Ref.~\cite{eisenstein_APL_1985}  in measurements of the magnetization oscillations in GaAs-AlGaAs heterojunction. Figure~\ref{fig:pudalov-hysteresis}  shows, that the hysteresis effect in chemical potential is observed both for  varying {$ \mu(H) =\int (\partial\mu/\partial H) dH$},
and electron density
 $ \mu(n) =\int (\partial\mu/\partial n) dn$.
The phenomenological interpretation of the observed hysteresis, suggested in \cite{pudalov_SSC_1984},  was confirmed  in subsequent studies,
however the microscopic origin of the effect for a long time remained a subject of debates.

The physical picture of nonstationary eddy currents excitation at first glance is simple  --
under magnetic field or electron density changes, in the quantum Hall effect regime,
a  relationship   $n=i\times n_H$ must hold between the number of electrons $n$ and flux quanta
$n_H= \Phi/\Phi_0$ (where $i$ -- integer).
This process requires   recharging current to flow in the 2D layer.
The Lorentz forces decline the charges coming into the 2D-layer, thereby causing eddy currents excitation.
The decay time of the eddy currents $\sim C/\sigma_{xx}$
tends to infinity for
$\sigma_{xx}= \rho_{xx}/\rho_{xy}^2 \rightarrow 0$.
In practice, however,
$ \rho_{xx}$ saturates below a certain temperature;
the corresponding saturation of the  decay time  allows to determine an important parameter --
the true value of the dissipative residual resistance of the  2D  system in the QHE regime,
undistorted by  contact effects.

The nonstationary recharging currents were studied
in the integer QHE
\cite{usher09, potts_JPCM_1996,
faulhaber_PRB_2005,
kershaw_NJP_2007}
and fractional QHE \cite{watts_PRL_1998} regimes. Beside macroscopic 2D  structures, the nonstationary
eddy currents  were  observed also in quantum dots in the QHE regime \cite{pioro_PRB_2006}.
The dynamics of eddy currents decay  was measured in many works
\cite{pudalov_SSC_1984, eisenstein_APL_1985, jones_SSC_1995, kershaw_NJP_2007}.
For
 GaAs-AlGaAs heterojunction in Ref.~\cite{eisenstein_APL_1985} the decay time was estimated as 300\,s at $T=400$\,mK.
In more detailed investigation of the eddy currents decay dynamics
performed at temperature  40\,mK in the QHE  $\nu = 4$ state
\cite{jones_SSC_1995},  the decay was found to be consistent with
exponential function whose
argument strongly varies with temperature, as expected
for the hopping-type conduction in the QHE regime.
However, for deeper resistance minima  $\nu=2,1$ a more complex  picture was found.

For the $\nu=2$ state,
eddy currents initially decay fast, with a characteristic time  $\tau_1\approx 40$\,s
which is related with a breakdown of the QHE by eddy currents.
Then, a slower process starts developing with a characteristic time  $\tau_2\approx 3.6$\,h.
Taking the  $\tau_2$ value
as an estimate of the true decay time in the low current regime, in Ref.~\cite{jones_SSC_1995}
an estimate was obtained for the residual resistance
at $T=40$mK: $\rho_{xx}^{min} \sim 10^{-14}$\,Ohm/$\Box$
for the $\nu=2$
and $10^{-11}$Ohm/$\Box$ for $\nu=4$ states.
Similar estimate was obtained in Ref.~\cite{pudalov_SSC_1984} for the  Si-MOS structure
($\nu=4$, $T=0.3$K): $\rho_{xx}^{min} \approx 10^{-11}$Ohm/$\Box$
and in  \cite{eisenstein_PRL_1985} for the  GaAs/AlGaAs heterojunction.

Thus, for a typical  capacitance of 1\,nF for a gated 2D structure, the characteristic recharging time
 $\tau= C/\sigma_{xx}=C\rho_{xy}^2/\rho_{xx}$
lies in the range  from $\sim 10^4$s
\cite{eisenstein_PRL_1985} to
$\sim 10^{10}$s  \cite{pudalov_SSC_1984, jones_SSC_1995}. These figures are cited here
for illustration of the time scale of the effect; of course, they depend on temperature,
on the relevant energy gap in the electron spectrum and on the Landau level broadening
 \cite{jones_SSC_1995}.

Consequently, the giant resistance drop $\rho_{xx}^{\rm min}/\rho_{xy}$ in the QHE regime
by a factor of  $\sim 10^{-14}- 10^{-17}$,
illustrates an empirical accuracy of reproducing the quantized resistance value in the Ohm standards
\cite{pudalov_JETP_1985, krasnopolin_IET_1987}.
Another practical application of the sharp peaks of nonstationary magnetic response under recharging
in the QHE regime is the control of homogeneity  of the 2D system.
Indeed, since
$\rho_{xx}$  raises exponentially sharp with deviation from the middle point between the Landau levels, the hysteresis effects occupy a narrow range in field or in density; the sharp response thereby uncovers the presence of domains with different concentration of delocalized states in the 2D  layer.

Already in the first  paper \cite{pudalov_SSC_1984} it was pointed that the eddy currents may flow locally, around the macroscopic localized areas in a smooth fluctuating potential landscape, or  along the real sample  edges, leading to a stored inductive or capacitive energy.  This issue was discussed in a number of papers   \cite{jones_SSC_1995, faulhaber_PRB_2005}, until  experimentally, using an electrometer with  submicron spatial resolution
 \cite{klaffs_PhysE_2007, huels_PRB_2004},
the profile of nonstationary current distribution was measured.
It was found, that indeed, the eddy current is concentrated mainly along the 2D system perimeter,
at a few micron distance from the   2D sample edges. This conclusion is consistent with magnitude of the eddy current  estimated from direct measurements using the torque magnetometer \cite{jones_SSC_1996}.

In spite of apparently exhausting answer from the experiments  with a spatial resolution,
the eddy current distribution, seems to be more complex \cite{jones_SSC_1996}.
The induced eddy currents circulate along the equipotential lines
in the presence of  potential fluctuations,
forming numerous current loops with various areas.
Each current loop  decays at its own rate, related with its capacitance and conductivity.
At the end of decay, for the remaining single loop,
the decay should occur exponentially with time. These arguments
 \cite{usher09} though plausible, however are not fully consistent with the fact, that the exponential law
was not observed in the experiments even after  24h.

Finally, the nonstationary currents were used as a valuable tool for contactless measurements of the
breakdown currents in the QHE regime, of the charge and current distribution in the sample in the QHE regime,
and also for estimating energy gaps in the electron spectrum \cite{jones_SSC_1996, matthews_PRB_2004, kavokin_SSC_2005, gething_IJMP_2004} -- the issues, interesting for physics and important for the  QHE metrology.

\subsubsection{Structure of the density of states in the QHE regime}

Measurements of the orbital electron magnetization were used in a number of studies for clarifying the energy structure of the density of states  $D(E)$  at the Landau levels, particularly,  in the gaps between the levels.
According to the semiclassical theory, for the ideal 2D gas with zero width of the Landau levels, $\Gamma=0$, magnetization should vary with field in saw-tooth fashion,  with the amplitude  $\mu_B^*$ per electron and with zero width of jumps in field \cite{ando-review, kukushkin_UFN_1988}. Approximately similar dependence was observed experimentally in high mobility GaAs/AlGaAs heterojunctions  \cite{wiegers_PRL_1997, schwarz_PRB_2002, wilde_PSS_2008}.

To account for disorder effect, in case of the isotropic elastic scattering  and ideal non-interacting electron gas, the density of states usually is described by Gaussian or Lorentzian function \cite{usher09, eisenstein_PRL_1985, kukushkin_UFN_1988}:
\begin{eqnarray}
D_{LL}^{G}(E) &\propto &\frac{1}{\pi l_B^2}\frac{1}{\sqrt{2\pi}\Gamma}\sum_{i=0}^\infty
\exp\left(-\frac{(E-E_N)^2}{2\Gamma^2}   \right) 
\nonumber\\
D_{LL}^{L}(E) &\propto & \frac{1}{\pi l_B^2} \sum_{i=0}^\infty \frac{\Gamma}{\left[(E-E_N)^2 +\Gamma^2\right]},
\end{eqnarray}
where $l_H=\sqrt{\hbar/eB}$ is  the magnetic length,
$E_N= (n+1/2)\hbar\omega_c$ -- energy of the  $N$-th Landau level, and $\Gamma$ --  level broadening.

It is well known however, that the experimentally measured density of states deviates from the Gaussian
dependence.
In many papers this deviation  is phenomenologically described by introducing a
background density of the in-gap states \cite{schwarz_PRB_2002} between the Landau levels:
 \beq
 D(E)= \zeta\frac{m^*}{\pi\hbar^2} +(1-\zeta) \frac{2eB}{\pi h}D_{LL}(E),
 \eeq
where the first term describes the energy independent density of states, and $\zeta$ --
is a fitting parameter.

In Refs.~\cite{zhu_PRB_2003, schwarz_PRB_2002, wilde_PRB_2006, kuntsevich_NatCom_2015}
the measured oscillations of the thermodynamic parameters for 2D electron system were compared with theory.  The shape of the measured quantum oscillations in \cite{pots_JPC_1996, tupikov_JETPL_2015} turned out to be described  in the best way using the Lorentzian distribution  with field independent  $\Gamma$, and by using $\zeta$  as an adjustable parameter. In contrast, in Ref.~\cite{wilde_PRB_2006, schwarz_PRB_2002}, the authors successfully  approximated the shape of   magnetization oscillations (and in Ref.~\cite{gornik_PRL_1985}-- shape of the electron specific heat) by using Gaussian distribution
with $\Gamma \propto \sqrt{B}$ and  with constant $\zeta$. Finally,
in Ref.~\cite{zhu_PRB_2003}, the oscillations were found to be equally well described with  Gaussian and Lorentzian distributions, with field independent $\Gamma$.

This apparent inconsistency of experimental results, in fact, finds an explanation in theoretical calculations for a smooth random potential \cite{sa-yakanit_PRB_1988, raikh_PRB_1993}; according to those in weak fields the Landau level width  must vary with field as $\sqrt{B}$,  whereas in strong field must saturates and become field independent.

The empirically determined non-zero width of  magnetization jumps  $\delta B$, i.e. the non-zero ``background'' density of states in the QHE regime is often attributed
to in-gap states, belonging to a separate reservoir of electron states, outside the  2D system. Within the framework of such approach, from the width of jump one can estimate the concentration of such states,
by describing it phenomenologically with the same parameter $\zeta$,
$n_{gap} = n\delta B/B$.
In particular, in Ref.~\cite{wiegers_PRL_1997, schwarz_PRB_2002} the authors estimated
 $n_{gap}/n\sim 2 - 3\%$ for $\nu=2$  in field of 12\,T.
However,
such huge  $\zeta =2- 3\%$ value
\cite{schwarz_PRB_2002, wiegers_PRL_1997},  and even
$\zeta =49\%$ \cite{zhu_PRL_2003},  we believe, make this hypothesis unphysical.

Quite similar idea of the existence of an electron reservoir
outside the 2D  system, where  electrons may enter  and quit, depending on the Fermi level position in the gap,
was discussed at the earlier stage of the QHE studies.
In order to test this assumption, in Ref.~\cite{pudalov-charge_JETPL_1984}
measurements were performed of the charge incoming the MOS structure. It was found experimentally,
that this charge coincides with the  charge of the 2D layer within experimental accuracy of  $< 2\%$;  in other words, the reservoirs of such huge capacity are missing in the Si-MOS structure.

In Refs.~\cite{gerhardts_PRB_1986, gudmundsson_PRB_1987} an attempt was performed to link the background density of states with statistical fluctuations of electron spatial distribution.
Another interpretation of the puzzling background density of states was suggested in \cite{semenchin_JETPL_1985, pudalov-quantosc_JETP_1985, wang_PRB_1988}:
the authors described the experimentally observed density of states using the Gaussian distribution whose
width $\Gamma(\nu)$ depends on the filling factor in oscillatory fashion.
Such interpretation is consistent with the concept of nonlinear screening and also with experimentally observed oscillations of the Landau level width \cite{semenchin_JETPL_1985, pudalov-quantosc_JETP_1985}.

\subsubsection{Renormalization of the oscillation amplitude of orbital magnetization by inter-electron interaction}
\label{renormalized osc}
As described above, the energy spectrum of  2D system
in quantizing perpendicular magnetic field $B$ consists of $\delta$-like discrete levels  (\ref{Landau_ladder})].
The magnetization  per electron in 2D system $\partial M/\partial n=-\partial E/\partial B$:
\beq
\frac{\partial M}{\partial n} = -\sum_N  \mu_B\left[ \frac{m}{m^*}\left(2N+1   \right) \pm \frac{1}{2}g^*\right].
\eeq
This relationship is fulfilled in all field intervals between the integer numbers of level fillings
($\nu=n/\Phi_0$ -- \textit{integer}, $\Phi_0=hc/e$-- flux quantum), where the magnetization experiences jump.
The amplitude of the jumps equals $2\mu_B (m/m^*)$  for cyclotron splittings (i.e. transitions  $N\rightarrow N\pm 1$), or   $g^*\mu_B$ -- for Zeeman splittings between levels with oppositely directed spins.

Non-zero temperature broadens the step-like changes of the filling function at the Fermi level,
that leads to broadening of the interval of the of jump-like changes in $\mu(H)$.
Disorder, in its turn,
causes broadening of the initially $\delta$-like Landau energy levels. As a result, both factors, temperature and disorder, cause diminishing of the jumps amplitude $\partial M/\partial n$.

When $e-e$ interaction is taken into account, the effective mass and {\em g}-factor  vary due to the Fermi-liquid renormalization, and the jumps amplitude must differ from the free electron value.
In quantizing magnetic field the renormalization (for the account of the so called
``inter Landau levels interaction''  or  ``level repulsion'')
leads to the enhancement in the jump amplitude. Such enhancement of the energy level splitting in the interacting  2D  electron system was observed experimentally and predicted theoretically \cite{macdonald_PRB_1986}.

\begin{figure}[h]
\begin{center}
\includegraphics[width=240pt]{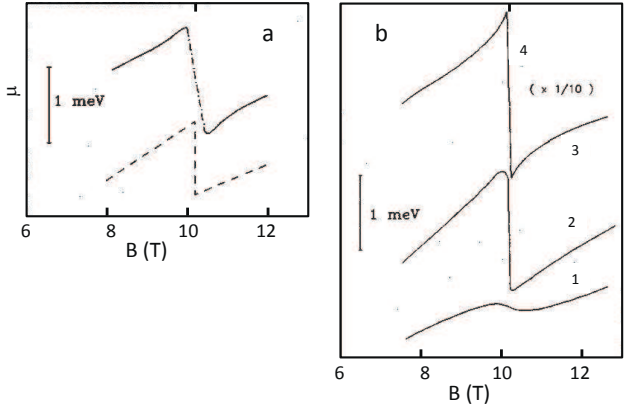}
\caption{a) Chemical potential (solid line) as a function of
field $B$  for the jump at $\nu=2$, from Ref.~\cite{krav-NDOS_physlettA_1990}
Temperature $T=1.3$\,K,  density of electrons $n=5\times 10^{11}$\,cm$^{-2}$.
Dashed curve shows  the calculated  $\mu(B)$ dependence for  2D system
of noninteracting electrons at $T=0$ and in the absence of disorder. b) Theoretical
$\mu(B)$ dependence
for two temperatures  $T$ and for the dimensionless Coulomb interaction contribution $\alpha$ \cite{krav-NDOS_physlettA_1990}:
$\alpha=0$, $T=0$ (curve 4), and $T=1.5$K, $\alpha=0$, 0.06 and 0.782 (curves 1, 2 и 3), respectively.
The latter value corresponds to the classical Coulomb interaction.}
\label{Fig:mu(H)-measured}
\end{center}
\end{figure}

Figure \ref{Fig:mu(H)-measured}a shows the measured chemical potential for
2D electron system in Si as a function of perpendicular magnetic field
 $B$ (the upper curve) \cite{krav-NDOS_physlettA_1990, krav_PRB_1990}. The sharp jump $\mu(B)$ at about 10\,T corresponds to the Fermi level transition from the 2nd to the 3rd energy level.
For the Fermi level location in the energy gap,
i.e. in the integer QHE regime, as was described above, resistance of the 2D system
decays exponentially strong,  its recharging under such conditions
is accompanied by eddy currents excitation, considered in section \ref{hysteresis}.
For this reason, the $\mu(B)$ behavior in Fig.~\ref{Fig:mu(H)-measured} in this range of fields  is schematically interpolated with a dash-dotted line.

Figure \ref{Fig:mu(H)-measured}b shows the  $\mu(B)$  dependence \cite{krav-NDOS_physlettA_1990}, calculated for the non-interacting  2D electron gas at $T=0$ in the absence of disorder, and also for a typical disorder-induced Landau level broadening. One can see, the slope of the measured dependence (i.e. magnetization per electron)
$\partial \mu/\partial B =-\partial M/\partial n$  for
 $\nu<2$ is about a factor of two greater than the maximum possible slope,
($\partial M/\partial n=\mu_B$), for the non-interacting electron gas.
The enhanced slope originates from the contribution of electron-electron interaction,
that also enters the inverse thermodynamic density of states (thermodynamic compressibility)
$(\partial n/\partial \mu)^{-1} $ and causes its negative value.
The latter effect was predicted  by Efros~\cite{efros88}  and experimentally observed in Refs.~\cite{krav-NDOS_PLA_1989, krav-NDOS_physlettA_1990, eisenstein_PRB_1992}.

\begin{figure}[h]
\begin{center}
\includegraphics[width=160pt]{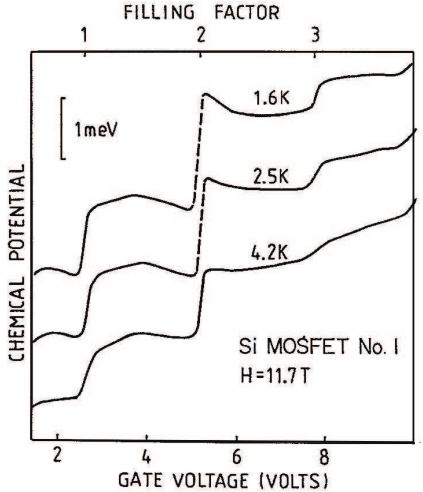}
\caption{Chemical potential dependence on electron density (controlled via the gate voltage $V_g$),
 measured for three temperatures. From Ref.~\cite{krav_PRB_1990}
}
\label{Fig:mu(Vg)-measured}
\end{center}
\end{figure}

Qualitatively, the  ``negative compressibility'' is clearly seen also in Fig.~\ref{Fig:mu(Vg)-measured},  where the chemical potential for two-dimensional electron system is shown versus electron concentration, measured in
constant magnetic field  \cite{krav_PRB_1990}. Instead of the anticipated (for non-interacting system)
step-like  $\mu(n)$ dependence with jumps
and with the related positive slope $d\mu/dn$ in the interval between them,
one can clearly see  intervals with $\mu/dn <0$. These wings with negative slope on  both sides of the integer fillings $\nu$
are the direct evidence for the negative contribution to the chemical potential due to the inter-electron interaction (i.e., negative compressibility).
The renormalized amplitude of the dHvA oscillations was measured in a number of works \cite{meinel_PRB_2001} and was found to be in a qualitative agreement with theory.

\subsection{Spin magnetization of electrons}
The problem of the electron spin magnetization measurements in 2D  systems became topical in 2000s,
in connection with investigations of the inter-electron correlation effects. The many body effects become progressively stronger in 2D systems under decreasing electron concentration; the latter, in its turn, became possible as a result of the  improvement in the quality of 2D structures.
Commonly, the inter-electron interaction is quantitatively characterized by a dimensionless ratio of the potential interaction energy and the kinetic Fermi energy, $r_s$ \cite{ando-review}.

 In order to study the effect of electron-electron correlations on the spin degree of freedom, numerous experiments were performed, using direct thermodynamic methods, as well as indirect
(i.e. based on theoretical models) transport methods;
their brief description and the major results are given below.

\subsubsection{Spin susceptibility renormalization, determined from oscillatory and monotonic transport}

\begin{figure}[h]
\begin{center}
\hspace{0.1in}\includegraphics[width=170pt]{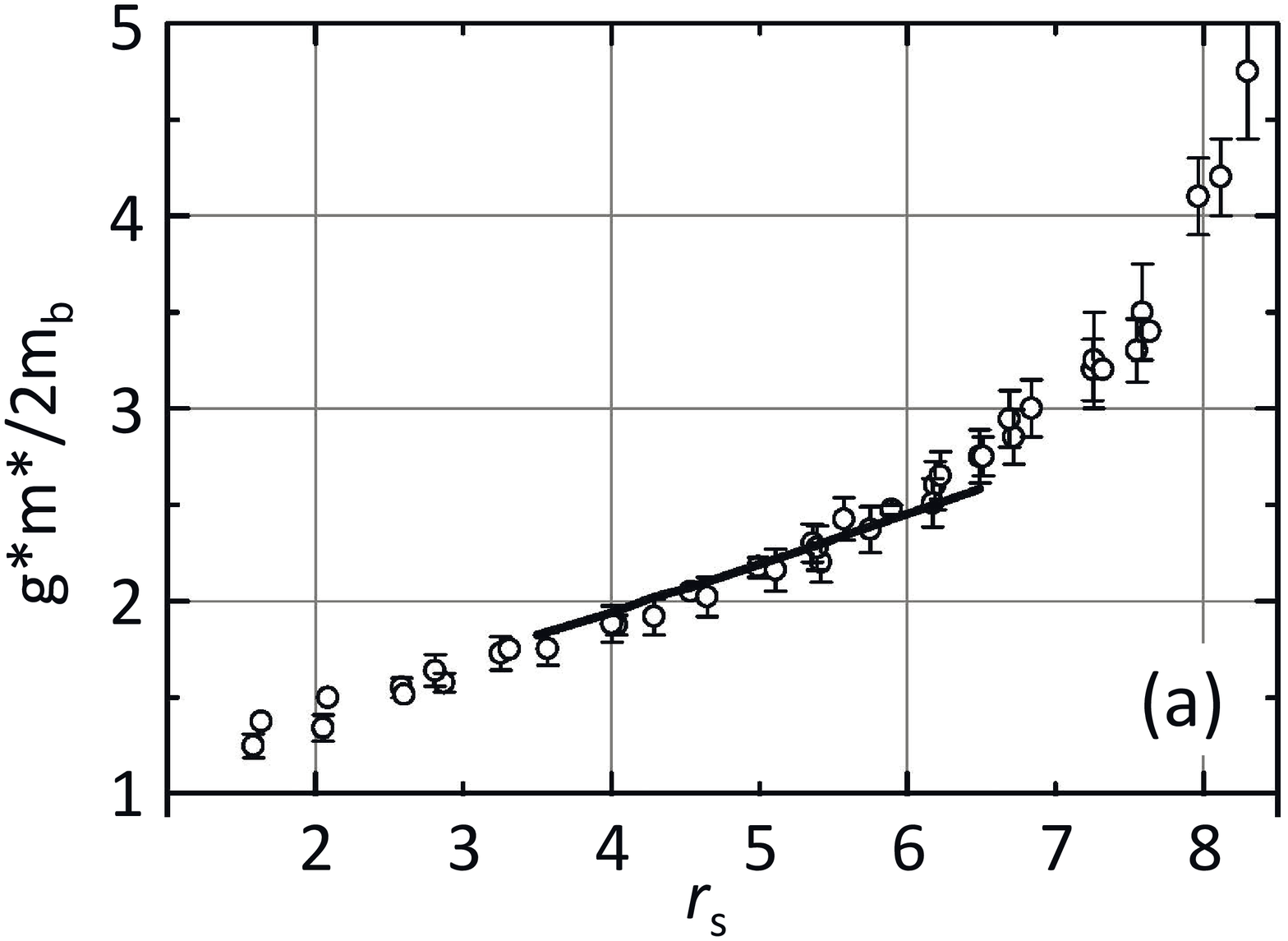}
\vspace{0.1in}
\includegraphics[width=150pt]{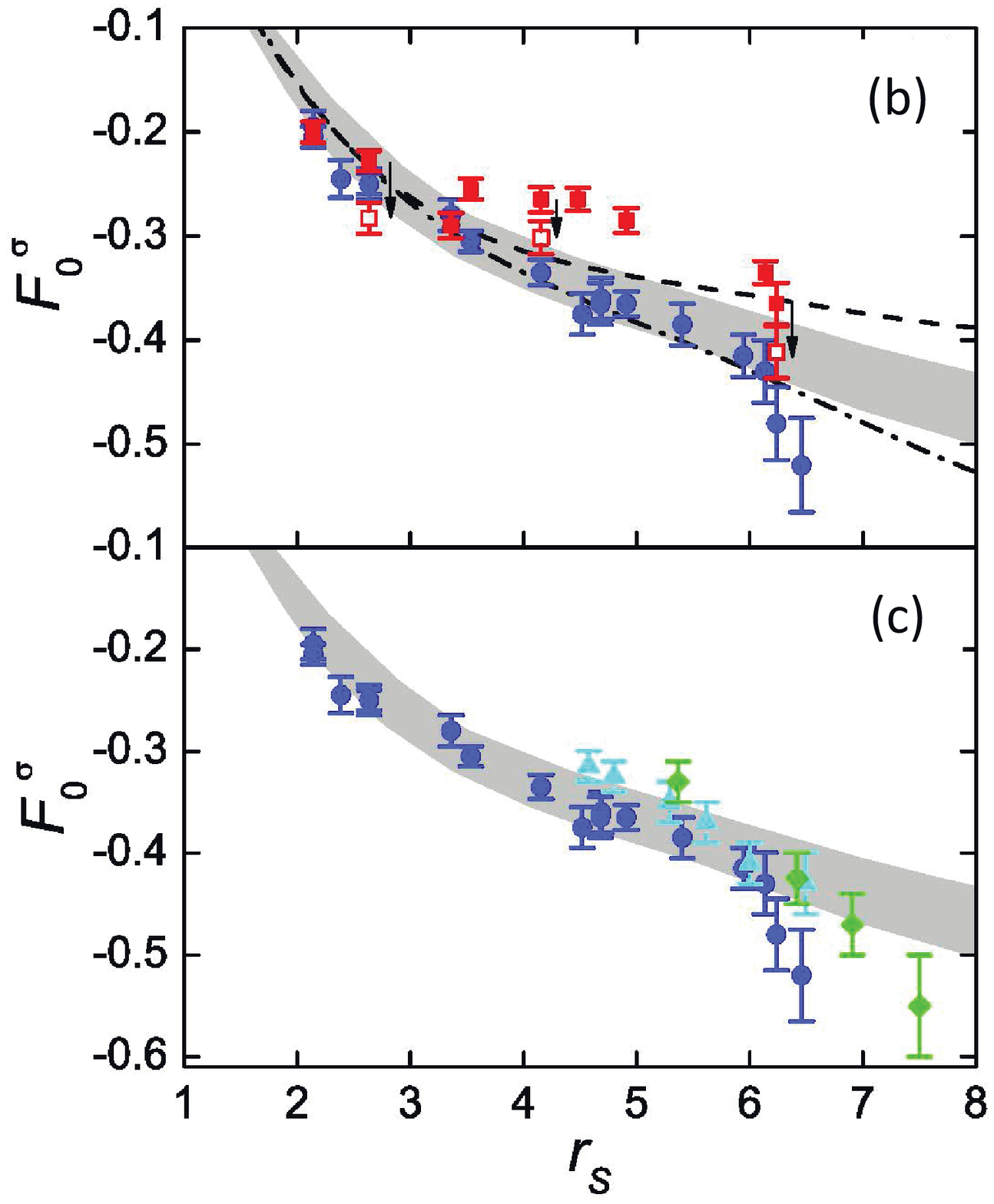}
\caption{(a) Spin susceptibility
$\chi^*/\chi_b$ dependence on  $r_s$, after \cite{gm_PRL_2002} (dots). Solid lines
-- data from  \cite{okamoto_PRL_1999}. (b)  $F_0^a(r_s)$ data,
obtained from  $\sigma(T)$ and $\sigma(B)$ fitting with theoretical dependence \cite{ZNA_PRB_2001}.
Dashed line -- $F_0^\sigma(r_s)$ values, extracted from the SdH \cite{gm_PRL_2002} measurements and using the LK theory \cite{LK_JETP_1955}; dash-dotted curve - empirical approach from \cite{gm_PRL_2002}. The shadowed corridor
represents the $F_0^\sigma(r_s)$ with experimental uncertainties,
obtained from SdH data fitting \cite{gm_PRL_2002} with theory Ref.~\cite{adamov_PRB_2006}.
(c)  $F_0^a$ values,  determined from $\sigma(T, B=0)$;
various symbols correspond to the data from \cite{klimov_PRB_2008},
as well as from \cite{shashkin_PRB_2002} and \cite{vitkalov_PRB_2003}, recalculated
as described in \cite{klimov_PRB_2008}.
}
\label{Fig:chi(rs)}
\end{center}
\end{figure}

Figure~\ref{Fig:chi(rs)}а  shows the main result,
summarizing measurements of $\chi^*/\chi_b\propto g^*m^*/2m_b$  for 2D  electron system in Si-MOS structures \cite{fang_PRB_1968, okamoto_PRL_1999, gm_PRL_2002, zhu_PRL_2003}.
One can see that, as a result of electron-electron interaction, the susceptibility   $\chi^* \propto g^*m^*$ increases monotonically with $r_s$ (i.e. with density lowering) by a factor of $\sim 5 $, though remains finite.

From the measurements  of  $\chi^*/\chi_b = g^*m^*/2m_b$ together with the  renormalized effective mass  $m^*(r_s)$,    one can extract the renormalized Lande $g^*$-factor  and, hence, to estimate the lowest order Fermi liquid coupling  parameter $F_0^\sigma$. The effective mass $m^*$ value may be found from temperature dependence of quantum oscillations. Fig.~\ref{Fig:chi(rs)}b shows the resulting  $F_0^\sigma(r_s)$ dependence, obtained from quantum oscillations; the results of Ref.~ \cite{zhu_PRL_2003} also agree with the data in Fig.~\ref{Fig:chi(rs)}b. As can be seen from Fig.~\ref{Fig:chi(rs)}c,
the $F_0^\sigma(r_s)$-values deduced from SdH oscillations reasonably agree with the
results, obtained from fitting  $\sigma(T, B=0)$ temperature dependence by the method, considered above
in Section  \ref{e-e-corrections}.

\begin{figure}[h]
\begin{center}
\includegraphics[width=230pt]{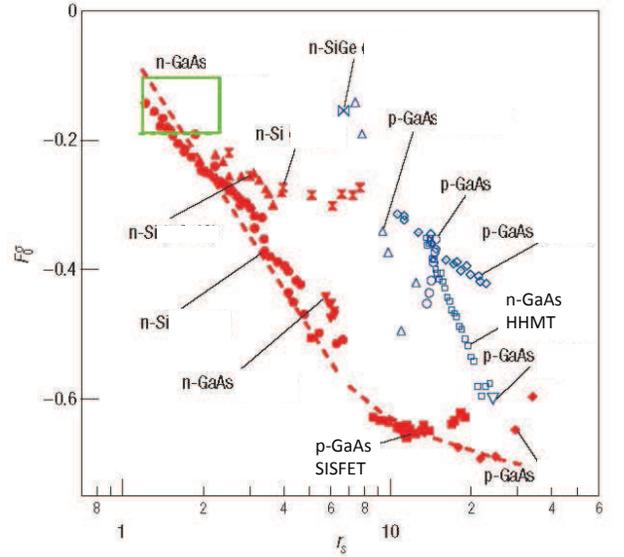}
\caption{Summary of $F_0^\sigma$-values, after Ref.~\cite{clarke_NatPhys_2007}. The red filled symbols are for the induced  2D systems, empty blue symbols  -  are for 2D systems of $n-$ and
$p-$type  in various materials. The green box surrounds the range of $F_0^\sigma$,
anticipated for 2D n-GaAs systems according to the theory of smooth potential screening \cite{clarke_NatPhys_2007}.
The dashed line is shown for clarity}
\label{Fig:clarke2007_Fig4}
\end{center}
\end{figure}

Finally,
Fig.~\ref{Fig:clarke2007_Fig4} from Ref.~\cite{clarke_NatPhys_2007} summarizes the results for 2D electron and hole systems; it demonstrates the impact of the character of disorder, clearly breaking the data into two groups, for the short-range and long-range  (as compared  with the Fermi wave length $\lambda_F\sim 100$\AA) potential fluctuations, which are described by theories  \cite{ZNA_PRB_2001} and \cite{gornyi_PRB_2004, gornyi_PRL_2003},  respectively.

For higher $r_s$ values, $F_0^\sigma$
tends to saturation at the level of $\sim -0.8$; as a result, the Stoner instability   expected for $F_0^\sigma=-1$, appears to be unattainable for all studied  2D material systems. Another cause of the attainability of the magnetic transition in single-phase 2D system is discussed in Section \ref{thermodynamic dM/dn}.

\subsubsection{Spin magnetization and susceptibility from thermodynamic measurements}
\label{thermodynamic dM/dn}
\begin{figure}[h]
\begin{center}
\includegraphics[width=170pt]{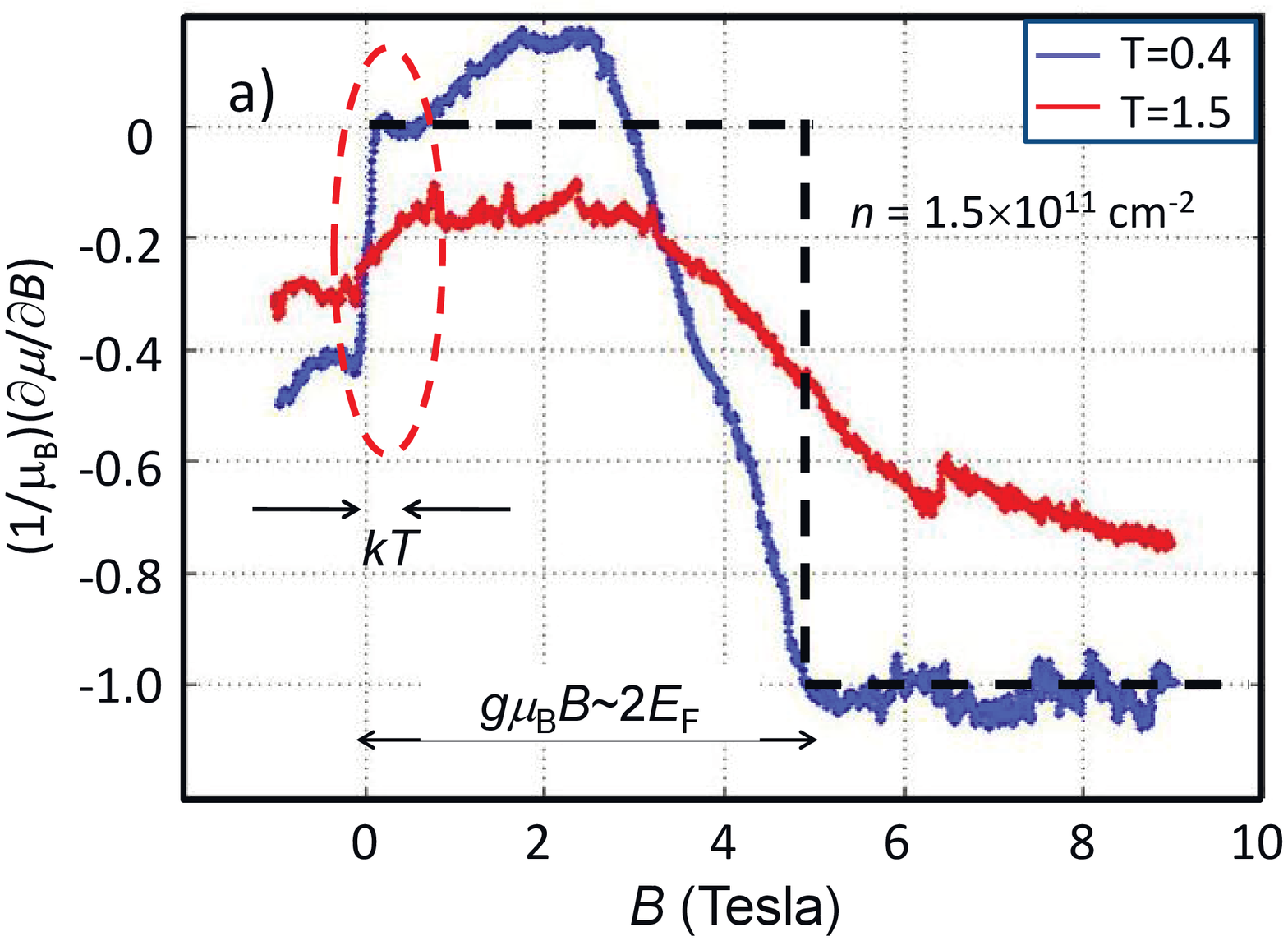}
\includegraphics[width=170pt]{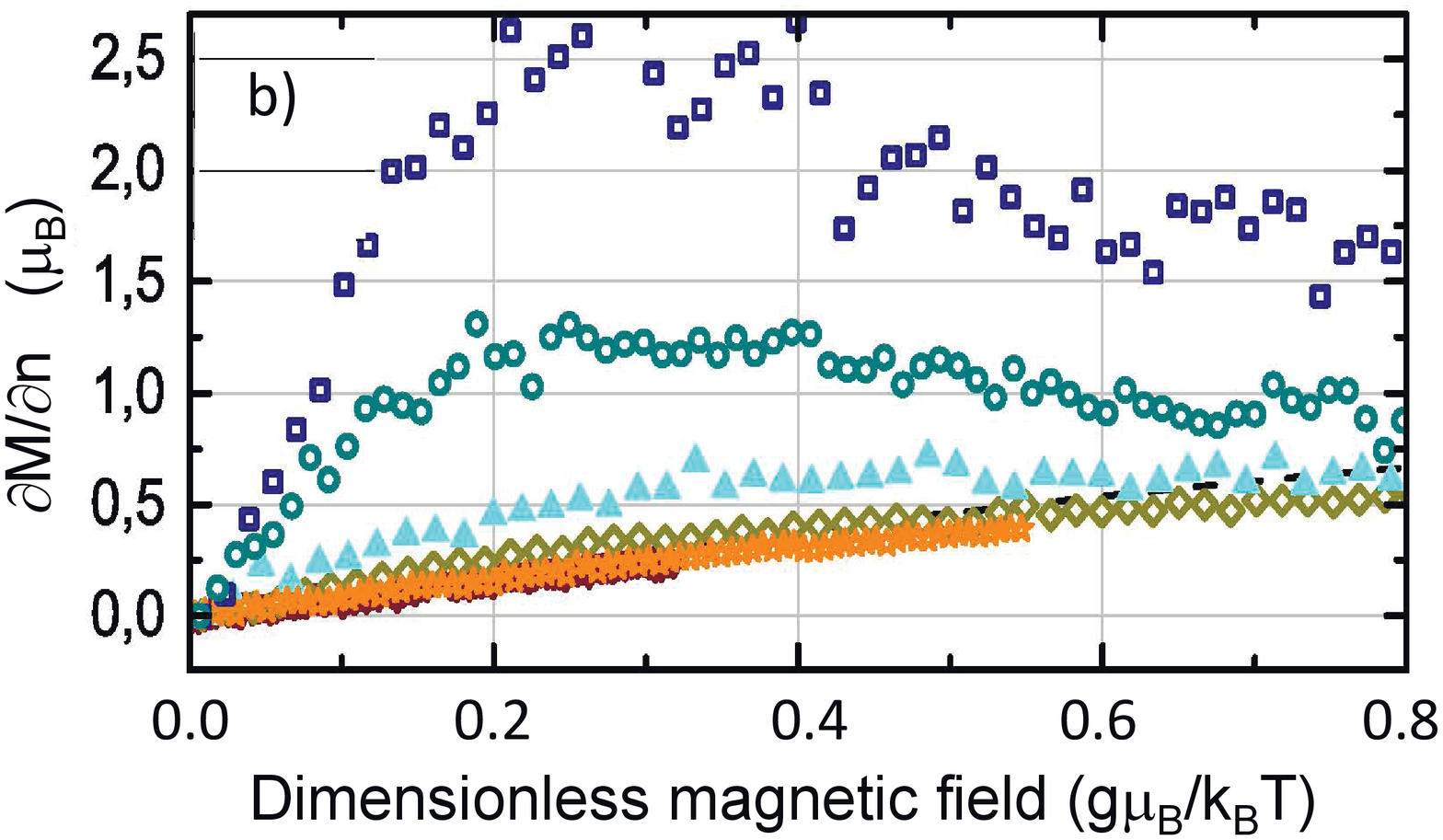}
\caption{(a) Typical $\partial\mu/\partial B=-\partial M/\partial n$ dependence on magnetic field for 2D electron  system  in  Si-MOS structure with a density $1.5\times 10^{11}$cm$^{-2}$.
Horizontal arrows mark the characteristic field ranges, corresponding to the normalized doubled Fermi energy,
 $\Delta B=(2E_F/g\mu_B)$,
and to the normalized temperature, $\delta B=(k_BT/g\mu_B)$. The dashes ellipse encloses a weak field region, zoomed in the lower panel. b): $dM/dn$ weak field dependence plotted versus normalized magnetic field $b=g\mu_B B/k_B T$ for carrier density  $0.5\times 10^{11}$cm$^{-2}$
at various temperatures ($T=0.8, 1.2, 1.8, 4.2, 7, 10, 24$K,  from top to bottom).}
\label{fig:dM/dn}
\end{center}
\end{figure}

The method of  $d\mu/dB$ thermodynamic measurements was described above in Section \ref{dmu/dB}.
Using this method, and by modulating the perpendicular $B_\perp$,  rather than parallel magnetic field, in Ref.~\cite{anissimova_PRL_2006} the renormalized  $g$-factor and cyclotron mass  $m^*$ was measured for 2D  electron system in Si;  evidently, the results include  orbital effects of inter-electron interaction. For probing purely spin effects, free of orbital contribution, measurements in Ref.~\cite{prus_PRB_2003, shashkin_PRL_2006} were performed in magnetic field aligned strictly parallel to the 2D plane. These results
taken in strong magnetic field  $B_\parallel \sim 2E_F/g\mu_B$ enable to detect features, expected for the full spin polarization (see Fig.~\ref{fig:dM/dn}).

In a partially polarized system, the electrons at the Fermi level have equal density of states for both spin projection and contribute  almost zero to the magnetization $dM/dn$.
Starting from the field of complete spin polarization, the $dM/dn$ value should sharply raise from 0 to $-\mu_B$,
as schematically shown by a bold dashed line in Fig.~\ref{fig:dM/dn}a.
A qualitatively similar behavior was observed in experiments \cite{prus_PRB_2003, shashkin_PRL_2006, dolgopolov_UFN_2019} and is shown in Fig.~\ref{fig:dM/dn}a.

Most of the measurements with this technique  was performed  in \cite{prus_PRB_2003, shashkin_PRL_2006}
in the regime of strong fields   $g\mu_B B\gg  k_BT$, which evidently ``cuts-off'' the $dM/dn$ temperature dependence. In the subsequent thermodynamic  measurements \cite{teneh_PRL_2012},
performed with improved sensitivity, a different behavior of $\partial M/\partial n$  was observed in weak fields  ($g\mu_B B<k_BT$), as shown in Fig.~\ref{fig:dM/dn}b.

Here, at high electron densities,  $\partial M/\partial n$ is negative  \cite{teneh_PRL_2012}, as expected for the Fermi-liquid because of
effective mass renormalization $\partial m^*/\partial n <0$.
At low densities  $\partial M/\partial n$ becomes positive and in all cases is much  greater than
 that expected for the Pauli spin susceptibility. When field increases (remaining all the way smaller than temperature),
$dM/dn$ sharply raises and, at low temperatures, exceeds  Bohr magneton more than by a factor of  two (Fig.~\ref{fig:dM/dn}b).

Such behavior of  $\partial M/\partial n (B)$ is reminiscent of the dependence, anticipated for the free spins,
$\partial M/\partial n = \mu_B \tanh(b)$, where $b = \mu_B B/k_BT \ll 1$  is the dimensionless magnetic field. However, the fact, that
$\partial M/\partial n$ exceeds Bohr magneton,
points at a ferromagnetic ordering of the electron spins. The magnetization curves
$\partial M/ \partial n$ (Fig.~\ref{fig:dM/dn}b) saturate in field of $b \approx 0.25$,  signalling that, the particles, which respond to the field modulation, have spin equal $\sim 1/2b \approx 2$, rather than (1/2).

Thus, the results of Ref.~\cite{teneh_PRL_2012} evidence for the emergence   of a two-phase state in 2D  system, consisting of paramagnetic Fermi liquid and ferromagnetic domains (so called ``spin droplets'') with a total spin  $\sim 2$, comprising $\gtrsim 4$ electrons. It seems likely, the formation of a two-phase state is more favorable, than transition to the uniform ferromagnetic state, that is in addition forbidden by the Mermin-Wagner theorem at $T\neq 0$.
 In the considered case, the easily orientable   ``nanomagnets'' remain persisting as the minority phase in the majority Fermi-liquid phase even though the dimensionless conductance  of the   2D system  $k_Fl >> 1$. Such conductance was commonly considered as a criterion of the well-defined Fermi-liquid state. We note, that the two-phase state often  occurs in interacting electron systems in the vicinity of phase transitions, expected for a uniform state \cite{gorbatsevich_1994, kornilov_PRB_2004, gerasimenko_PRB_2014}.

\section{Conclusion}
Magnetic properties measurements of non-magnetic or weakly-magnetic materials
always represented a topical task, relevant for both practical material applications, and physical studies.  The doubtless advantage of magnetometry  is related with thermodynamic character of measurements,
 that in many cases, provides related simple and reliable interpretation of the results.
Experimental methods of the magnetic measurements are continuously improved, mostly since the end of the previous century. This review considers various methods of magnetometry  and their evolution in the last 50 years.
As a result of their development, dozens of outstanding laboratory magnetometer designs appeared,  followed by a large number of commercially available magnetometers and susceptometers.

The demand in magnetic measurements raised sharply in the beginning of 1970s, related with discovery and intensive studies of low-dimensional systems of electrons in the semiconductor structures  \cite{ando-review}  and in organic crystals  \cite{lebed-book, kornilov_PRB_2004, kornilov_PRB_2002}.
Low -dimensional electron systems manifest a rich novel physics in strong magnetic fields. Beside the traditional transport and optical measurements, their studies  require also thermodynamic, and particularly, magnetic measurements. Investigations of orbital magnetization of low-dimensional electron systems and nanostructures with low number of electrons has required improving traditional designs and developing novel methods for magnetic measurements. Along with discovery and studies of the integer and fractional quantum Hall effects, simultaneously  performed magnetic measurements with 2D electron systems has led to a deeper understanding of the origin of these effects, properties of novel quasiparticles, describing the fractional charge states, composite quasiparticles, consisting of electrons and flux quanta, and collective spin excitations in the electron systems.

In the beginning of the  21st century, the problem of a weaker effects of electron spin magnetization  came to the forefront. This is related with the topical  problem of understanding properties of strongly correlated electron systems,  searching novel states of electron matter, studying effects of spin ordering and their interplay  with superconducting paring, as well as with application in spintronics  and quantum computations.

And at last, in recent years there were developed new methods of magnetometry  with spatial and temporal resolution. Local probing uses such tools as scanning  magnetometers based on the  NV-centers, SQUID-magnetometers, scanning Hall magnetometers, and scanning atomic force microscopes. The time resolved magnetometry  enables studying magnetization dynamics during  relaxation of the system between two quantum states. These methods have great perspectives, because they are suited to magnetic measurements with more and more popular nanomaterials, nanostructures of topologically non-trivial matter, and optically controlled matter. The magnetometry methods with nm-spatial resolution and temporal resolution are now quickly developing,  adapting to novel tasks and will promote novel discoveries, and  accumulation of novel knowledge, particularly in  such topical areas as studies of the quantum topological effects, novel quasiparticles  (including, e.g. Majorana fermions), living cells, microorganisms and neuro-systems. Scanning magnetic local microscopy here suggests a  unique possibility of non-invasive probing and visualization of the structure and dynamics of nano-objects.

\section{Acknowledgements}
The authors is grateful to  M.E. Gershenson, E.M. Dizhur, G. Bauer, G. Brunthaler,
 N. Klimov, H. Kojima, S.V. Kravchenko,
A.Yu. Kuntsevich,  L.A. Morgun, M. Reznikov, D. Rinberg, S.G. Semenchnisky,  N. Teneh,  and  V.S. Edel'man,
for fruitful collaboration in developing experimental methods, performing  measurements,  discussing the results, and writing the original papers.
Financial support from RFBR \#18-02-01013 is acknowledged.

\end{document}